%% file: minDM.tex
\documentclass{article}

\usepackage{authblk,amsmath,amssymb,graphicx,cite,hyperref}
\usepackage{geometry}
\usepackage[utf8]{inputenc}
\DeclareUnicodeCharacter{2009}{\,}

\usepackage{tikz}
\usetikzlibrary{positioning,arrows,patterns}
\usetikzlibrary{decorations.pathmorphing}
\usetikzlibrary{decorations.markings}
\usetikzlibrary{calc}
\tikzset{
    photon/.style={decorate, decoration={snake}, draw=black, thick},
    fermionnoarrow/.style={draw=black, postaction={decorate}, thick},
    scalar/.style={draw=black, postaction={decorate}, thick, dashed},
    fermion/.style={draw=black, postaction={decorate},decoration={markings,mark=at position .55 with {\arrow{>}}}, thick},
    gluon/.style={decorate, draw=black, decoration={coil,amplitude=4pt, segment length=5pt}, thick},
    vertex/.style={draw,shape=circle,fill=black,minimum size=3pt,inner sep=0pt}
}

\newcommand{\modeqref}[1]{eq.~\eqref{#1}}
\newcommand{\modeqsref}[1]{eqs.~\eqref{#1}}

\newcommand{\refcite}[1]{ref.~\cite{#1}}
\newcommand{\refscite}[1]{refs.~\cite{#1}}

\newcommand{\figref}[1]{fig.~\ref{#1}}
\newcommand{\Figref}[1]{Fig.~\ref{#1}}
\newcommand{\figrefs}[1]{figs.~\ref{#1}}
\newcommand{\tabref}[1]{table~\ref{#1}}

\newcommand{\secref}[1]{section~\ref{#1}}
\newcommand{\secrefs}[1]{sections~\ref{#1}}

\newcommand{\set}[1]{\mathbb{#1}}

\graphicspath{{./}{./Figures/}}

\begin{document}

\title{Electroweak Multiplet Dark Matter at Future Lepton Colliders}
\author[1]{Kenji Kadota}
\author[1]{Andrew Spray}
\affil[1]{Center for Theoretical Physics of the Universe, Institute for Basic Science (IBS), Daejeon, 34126, Korea}
\maketitle

\begin{abstract}
	An electroweak multiplet stable due to a new global symmetry is a simple and well-motivated candidate for thermal dark matter.  We study how direct searches at a future linear collider, such as the proposed CLIC, can constrain scalar and fermion triplets, quintets and septets, as well as a fermion doublet.  The phenomenology is highly sensitive to charged state lifetimes and thus the mass splitting between the members of the multiplet.  We include both radiative corrections and the effect of non-renormalisable operators on this splitting.  In order to explore the full range of charged state lifetimes, we consider signals including long-lived charged particles, disappearing tracks, and monophotons.  By combining the different searches we find discovery and exclusion contours in the mass-lifetime plane.  In particular, when the mass splitting is generated purely through radiative corrections, we can exclude the pure-Higgsino doublet below 310\;GeV, the pure-wino triplet below 775\;GeV, and the minimal dark matter fermion quintet below 1025\;GeV.  The scenario where the thermal relic abundance of a Higgsino accounts for the whole dark matter of the Universe can be excluded  if the mass splitting between the charged and neutral states is less than 230\;MeV.  Finally, we discuss possible improvements to these limits by using associated hard leptons to idenify the soft visible decay products of the charged members of the dark matter multiplet.
\end{abstract}

\section{Introduction}

\input{./Files/Introduction}

\section{Models}\label{sec:models}

\input{./Files/Models}

\section{Signal and background event generation}\label{sec:evgen}

\input{./Files/evgen}

\section{Long lived charged particles}\label{sec:llcp}

\input{./Files/Muons}

\section{Disappearing tracks}\label{sec:dtrack}

\input{./Files/Tracks}

\section{Monophotons}\label{sec:monophotons}

\input{./Files/Mono-Photons}

\section{Combined limits in the mass-lifetime plane}\label{sec:comb}

\input{./Files/comb}

\section{Discussion}\label{sec:pions}

\input{./Files/Pions}

\section{Conclusions}\label{sec:conc}

\input{./Files/conc}

\section*{Acknowledgements}

We thank Filippo Sala and Marco Cirelli for collaborating in the early stages of this project.  This work was supported by IBS under the project code, IBS-R018-D1.

\bibliography{minDM}{}
\bibliographystyle{JHEP}

\end{document}

%% file: Files/Introduction.tex
The dark matter (DM) problem remains perhaps the most compelling sign for the need for physics beyond the Standard Model (SM).  While there exists ever-growing support from astrophysical observables for its existence, there remains no unambiguous direct signal at terrestrial experiments despite substantial recent experimental progress.  In this environment we are obliged to consider all possible avenues of exploration.  In particular, we should ask how and to what extent experiments currently under design can illuminate the nature of DM.

Thermal freeze-out remains a popular and compelling explanation for the observed DM abundance.  It is insensitive to the cosmological initial conditions, generic for stable particles, and predicted by models such as supersymmetry.  One of the simplest examples is a new scalar or fermion electroweak multiplet, with an appropriately chosen hypercharge to ensure a neutral component, and stable due to a new symmetry.  The well-studied Higgsino and wino of supersymmetry are among this class.  In the decoupling limit where all other states are heavy and the hypercharge is zero, these models are most strongly probed by indirect cosmic ray searches~\cite{1507.05519,1507.05536}.  Unfortunately these limits will always contain systematic uncertainties arising from the DM density distribution within the galaxy, the cosmic ray propagation model, and other sources.  Searches using terrestrial experiments remain important to check and corroborate the limits that exist.

In this work, we will consider collider searches for direct production of electroweak multiplet DM.  In the absence of new coloured states, limits from lepton colliders are generally superior to those from hadron machines.  A number of proposals have been made for future $e^+e^-$ experiments, including the International Linear Collider (ILC)~\cite{1306.6327} as well as future circular colliders~\cite{1305.6498}.  However, the reach of direct searches is limited to half the centre of mass energy, which motivates us to consider the proposal with the largest $\sqrt{s} = 3$\;TeV, the Compact Linear Collider (CLIC)~\cite{1202.5940}.  Because CLIC is a more speculative proposal, we restrict ourselves to robust signals based on energetic simple final states that are unlikely to vary much as the experimental design changes.  Our goal is to provide estimates of the discovery potential and exclusion reach that are conservative but comparable to the final sensitivity.  We do also discuss some more speculative possibilities that might cover regions of parameter space that are otherwise poorly constrained.

The models we consider are defined by two parameters: the overall multiplet mass $m_\chi$; and the splitting between the DM and the singly-charged member $\psi^+$ of the electroweak multiplet $\Delta m_1$ (which, for real multiplets, is bounded $\lesssim 1$\;GeV).  As DM production is an electroweak process, the signal cross section is fixed by $m_\chi$, and charged states are dominantly produced.  The phenomenology is set by the charged state lifetimes, which in turn are determined by the mass splitting.  This leads us to consider three distinct phases, in order of increasing $\Delta m_1$:
\begin{itemize}
	\item When $\psi^+$ is collider stable, the relevant searches are for long-lived charged particles depositing energy in the muon chambers.  We find that due to small backgrounds, searches at linear colliders are very strong, excluding $m_\chi$ up to half the centre of mass energy.
	\item As the lifetime of $\psi^+$ increases, it will decay within the detector volume.  It will then leave charged tracks that terminate before reaching the muon chamber.  This leads to a `disappearing tracks' signal commonly associated with winos; the limits here are weakened by an uncertainty in the background, but can still be quite strong.  
	\item When $\psi^+$ decays promptly, it is the least constrained at CLIC.  Identifying the soft visible decay products is challenging due to coincident $\gamma\gamma \to $~hadrons activity.  Ignoring them motivates a monophoton search, but the reach is limited by the large $e^+e^-\to\nu\nu\gamma$ background.
\end{itemize}
By combining all three cases, we can exclude $\Delta m_1 \lesssim 100$\;MeV for almost any multiplet up to the maximum mass that can be produced ($m_\chi = 1.5$\;TeV).  The constraints for $\Delta m_1 \lesssim 200$\;MeV are also generally strong.  At greater mass splittings, only large multiplets with enhanced production cross sections can be easily tested.

Dark matter at future colliders is an active area of study.  The prospects from indirect searches at CLIC are discussed in \refcite{1810.10993}, and at other proposed lepton colliders in \refscite{1504.03402,1604.07536}.  For a discussion of winos at a 5\;TeV lepton collider, see \refcite{Chattopadhyay:2006xb}, and for limits on electroweak multiplet DM at future hadron colliders see \refscite{1404.0682,1703.05327,1805.00015}.  Searches at future lepton colliders for models with two different electroweak multiplets were considered in \refscite{1611.02186,1705.07921,1707.03094,1711.05622}.


The outline of this paper is as follows.  We first define our dark matter models in \secref{sec:models}, in particular the mass splittings among members of the multiplet and the lifetimes of the charged states.  We outline general aspects of our event generation in \secref{sec:evgen}.  We then consider the limits that arise from different signals in the following sections: long-lived charged particles in \secref{sec:llcp}; disappearing tracks in \secref{sec:dtrack}; and monophotons in \secref{sec:monophotons}.  We combine all limits in the mass-lifetime plane in \secref{sec:comb}.  Possible avenues for improving the limits we find using the soft decay products of the charged states are discussed in \secref{sec:pions}.  Finally we conclude in \secref{sec:conc}.

%% file: Files/Models.tex
We assume that dark matter $\chi$ consists of the neutral component of a fermion or scalar electroweak multiplet $\psi$.  Stability is enforced by the presence of an unbroken global $\set{Z}_2$ symmetry under which all SM fields transform trivially.  If the multiplet has non-zero hypercharge, $\chi$ will be complex and have unsuppressed couplings to the $Z$, resulting in severe direct detection constraints.  Indeed, such models are generally excluded~\cite{0710.1668,1711.09912} unless there is a mass splitting of $\chi$ into two real fields, such that the $Z$ coupling becomes inelastic with $\delta m \gtrsim 140$\,keV~\cite{TuckerSmith:2001hy,1608.02662}.  We therefore focus on hypercharge-zero multiplets, with the sole exception of a fermion doublet with $Y$ equal to one-half, \emph{i.e.} the same SM quantum numbers as the Higgsino.  To be concrete, we consider scalar and fermion triplets, quintets and septets.  The fermion triplet (quintet) is similar to a pure Wino (Minimal Dark Matter~\cite{hep-ph/0512090}), such that the collider limits we derive below apply in those cases also.

\begin{figure}
	\centering
	\begin{tikzpicture}[node distance=0.75cm and 0.75cm]
		\coordinate (v1);
		\coordinate[right = of v1, label=below:$Z/\gamma$] (v2);
		\coordinate[right = of v2] (v3);
		\coordinate[above left = of v1, label=above left:$e^-$] (i1);
		\coordinate[below left = of v1, label=below left:$e^+$] (i2);
		\coordinate[above right = of v3, label=above right:{$\psi^{q-}$}] (o1);
		\coordinate[below right = of v3, label=below right:{$\psi^{q+}$}] (o2);
		\draw[fermion] (i1) -- (v1);
		\draw[fermion] (v1) -- (i2);
		\draw[photon] (v1) -- (v3);
		\draw[fermionnoarrow] (o2) -- (v3);
		\draw[fermionnoarrow] (v3) -- (o1);
	\end{tikzpicture}\qquad\qquad
	\begin{tikzpicture}[node distance=0.75cm and 0.75cm]
		\coordinate (v1);
		\coordinate[right = of v1] (v2);
		\coordinate[right = of v2] (v3);
		\coordinate[above left = of v1, label=above left:$e^-$] (i1);
		\coordinate[below left = of v1, label=below left:$e^+$] (i2);
		\coordinate[above right = of v3, label=above right:{$\psi^{q-}$}] (o1);
		\coordinate[below right = of v3, label=below right:{$\psi^{q+}$}] (o2);
		\coordinate[left = of o1, label=above right:{$Z$}] (o3);
		\draw[fermion] (i1) -- (v1);
		\draw[fermion] (v1) -- (i2);
		\draw[photon] (v1) -- (v2);
		\draw[photon] (v2) -- (o3);
		\draw[scalar] (v2) -- (v3);
		\draw[fermionnoarrow] (o2) -- (v3);
		\draw[fermionnoarrow] (v3) -- (o1);
	\end{tikzpicture}
	\caption{Example production channels of electroweak multiplets at CLIC.  (Left): production through gauge couplings.  (Right): Production through Higgs portal, allowed for scalar dark matter but typically negligible.  See the text for more details.}\label{fig:prod}
\end{figure}
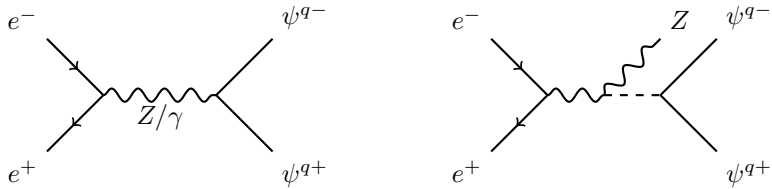

Our interest is in the collider phenomenology of direct production of the $\psi$ multiplet, \emph{e.g\@.} when there are no other kinematically accessible new states.  The gauge couplings of the multiplet are the natural production mechanism, and for fermions the only possible renormalisable coupling after integrating out all other states.  The choice of zero hypercharge to avoid direct detection limits implies that there is no tree-level production of the $\chi\chi$ state.  Instead, the dominant channel into the dark sector is $SM\,\overline{SM} \to Z^\ast/\gamma^\ast \to \psi^{q+}\psi^{q-}$, where $q$ is the charge, followed by decay of the charged states; see the left-side of \figref{fig:prod}.  We will also consider related production channels, in particular those with an additional final state photon.  The detector signals will be sensitive to the decay modes and lifetime of the charged states in the multiplet, which in turn are determined by the mass splitting between them and the DM.

In the simpler case of fermionic DM, there is only a single new renormalisable parameter before electroweak symmetry breaking, namely the mass $m_\psi$.  The mass splitting between the different components of $\psi$ is then entirely determined by radiative corrections.  The neutral component $\chi$ is lightest, and the splitting of the charged components is given by~\cite{hep-ph/0512090}
\begin{equation}
	\Delta m_q \equiv m_{\psi^{q+}} - m_\chi \simeq 166 \, q^2 \, \text{MeV.} 
\label{eq:raddm}\end{equation}
In this work, we are more general and consider the mass splitting to be a free parameter.  From the low-energy point of view, we can achieve this by adding higher-dimensional terms.  For the complex doublet, the leading contribution comes from the dimension-5 operator
\begin{equation}
	\mathcal{L} \supset \frac{c_{5\psi}}{\Lambda} \, (H^\dagger T^a_H H) \, \bar{\psi} T^a_\psi \psi \,,
\label{eq:d5split}\end{equation}
where $T^a_i$ are the $SU(2)_L$ generators for the representation of $i$.  This will contribute a mass splitting
\begin{equation}
	\Delta m_q \sim c_{5\psi} \, (q - Y) \, \frac{v_h^2}{2\Lambda} = 153 \,(q - Y) \; \text{MeV} \, \biggl( \frac{100 \, \text{TeV}}{\Lambda/c_{5\psi}} \biggr) \,,
\label{eq:dmlinear}\end{equation}
with $Y$ the hypercharge of the multiplet.  This can  increase or decrease the mass splitting depending on the sign of the Wilson coefficient $c_{5\psi}$, and can easily dominate the radiative mixing of \modeqref{eq:raddm} for not-too-large $\Lambda$.  However, \modeqref{eq:dmlinear} vanishes when $\psi$ is a Majorana fermion; the adjoint combination of two real representations of $SU(2)$ is antisymmetric, but the spinor contraction $\bar{\psi}\psi$ is symmetric.  The leading contribution instead arises at dimension 7~\cite{Ibe:2012sx}:
\begin{align}
	\mathcal{L} & \supset \frac{c_{7\psi}}{\Lambda^3} \, (H^\dagger T^a_H H)  \, (H^\dagger T^b_H H) \, \bar{\psi} T^a_\psi T^b_\psi \psi \,,\label{eq:d7split} \\
	\Delta m_q & \sim c_{7\psi} \, q^2 \frac{v_h^4}{4\Lambda^3} = 69 \, q^2 \, \text{MeV} \times c_{7\psi} \biggl( \frac{1.5\;\text{TeV}}{\Lambda} \biggr)^3 \,.
\end{align}
If we interpret $\Lambda$ as a physical mass scale associated with additional matter, then our assumptions require $\Lambda \geq 1.5$\;TeV so they are not produced at CLIC.  We see that for this mass splitting to be larger than the radiative piece, one requires moderately large values of the Wilson coefficient, $c_{7\psi} \gtrsim 3$, suggesting a strongly coupled UV completion.  If we take the na\"\i ve dimensional analysis limit $\lvert c_{7\psi} \rvert < 4\pi$, the mass splitting (including radiative piece) is bound by $\Delta m_1 \lesssim 1$\;GeV.  Larger mass splittings require either extra light states or making $\psi$ a Dirac fermion\footnote{This would be necessary if the symmetry stabilising the dark matter is anything other than $\set{Z}_2$.  It would also increase all collider production cross sections by two, increasing the limits we find later.}.

For the fermion doublet and triplet, we need only specify the DM mass and the mass splitting $\Delta m_1$ to define the model.  The details of precisely \emph{how} we generate the mass splitting are not important.  For the higher multiplets, we must also specify the mass splittings for the higher charged states.  Production is proportional to $q^2$ so these states typically dominate collider processes.  We adopt the minimal choice of using \modeqref{eq:d7split} with $\Delta m_1$ as the input; then the mass splittings between adjacent charged states are 
\begin{subequations}
	\begin{align}
		m_{\psi^{2+}} - m_{\psi^+} & = 3 \, \Delta m_1 \,, \\
		m_{\psi^{3+}} - m_{\psi^{2+}} & = 5 \, \Delta m_1 \,.
	\end{align}
\label{eq:dmhighq}\end{subequations}
The higher charged states decay more rapidly than the singly-charged one, but the mass splittings remain $\mathcal{O} ($GeV).  As we discuss in more detail later, reconstructing such soft decay products is experimentally challenging.  The collider phenomenology almost entirely determined by the lifetime of the longest-lived state, which is always the singly-charged state $\psi^+$.

Scalar dark matter models are marginally more complex, as they involve renormalisable scalar quartic couplings:
\begin{equation}
	V(H,\phi) \subset \frac{1}{2} \, \lambda_\psi \lvert \psi \rvert^4 + \frac{1}{2} \, \lambda_{h\psi} \, \lvert \psi \rvert^2 \, H^\dagger H \,.
\label{eq:phipot}\end{equation}
The first term is a self-interaction generally irrelevant to DM physics; the second the well-known Higgs portal coupling.  The latter can potentially lead to additional collider signals which would complicate our phenomenology.  At an $e^+e^-$ collider, the tiny electron Yukawa means that (virtual) Higgses arise through vector boson fusion and/or Higgsstrahlung.  As can be seen from \figref{fig:prod}, dark sector production through the Higgs portal will involve at least one additional final state particle and the same number of electroweak couplings as compared to through gauge bosons alone, and hence are suppressed by at least $\lambda_{h\psi}^2/(4\pi)^2$.  In comparison, direct detection searches impose the constraint $\lambda_{h\psi} \lesssim 0.1$\,--\,0.01 for $m_\chi$ in the range 100\,--\,1000\;GeV~\cite{1701.08134}, so production through the $\psi$ gauge coupling always dominates.

In addition to the couplings of \modeqref{eq:phipot}, we might expect two additional terms involving $SU(2)$ generators:
\begin{equation}
	\frac{1}{2} \, \lambda_\psi' \, (\psi^T T^a_\psi \psi)^2 + \frac{1}{2} \, \lambda_{h\psi}' \, (\psi^T T^a_\psi \psi) \, (H^\dagger T^a_H H) \,.
\end{equation}
The second term, in particular, would contribute to the mass splitting between the elements of the multiplet.  However, while these terms exist for complex scalars, for real scalars they vanish (again, because the adjoint combination of two real representations is antisymmetric while $\psi^T\psi$ is symmetric).  Instead, the leading contribution to the mass splitting comes at dimension-6,
\begin{align}
	V(H,\phi) & \subset \frac{c_{6\psi}}{2\Lambda^2} \, (H^\dagger T^a_H H) \, (H^\dagger T^b_H H) \, \psi^T T^a_\psi T^b_\psi \psi \,, \label{eq:d6split} \\
	\Delta m_q & \sim c_{6\psi} \, q^2 \, \frac{v_h^4}{8 \Lambda^2 m_\chi} \sim 104 \, q^2 \, \text{MeV} \times c_{6\psi} \, \biggl(  \frac{500\;\text{GeV}}{m_\chi} \biggr) \biggl(\frac{1.5\;\text{TeV}}{\Lambda} \biggr)^2 \,.
\end{align}
We can achieve larger mass splittings for the scalar multiplet than the fermion.  However, we still have the rough bound $\Delta m_1 \lesssim 1$\;GeV absent additional light states or complex DM.  Most importantly, the mass splitting has the same scaling with charge as the radiative piece~\eqref{eq:raddm}, so that the relations of \modeqref{eq:dmhighq} apply for scalar as well as fermion dark matter.

\begin{figure}
	\centering
	\includegraphics[width=0.45\textwidth]{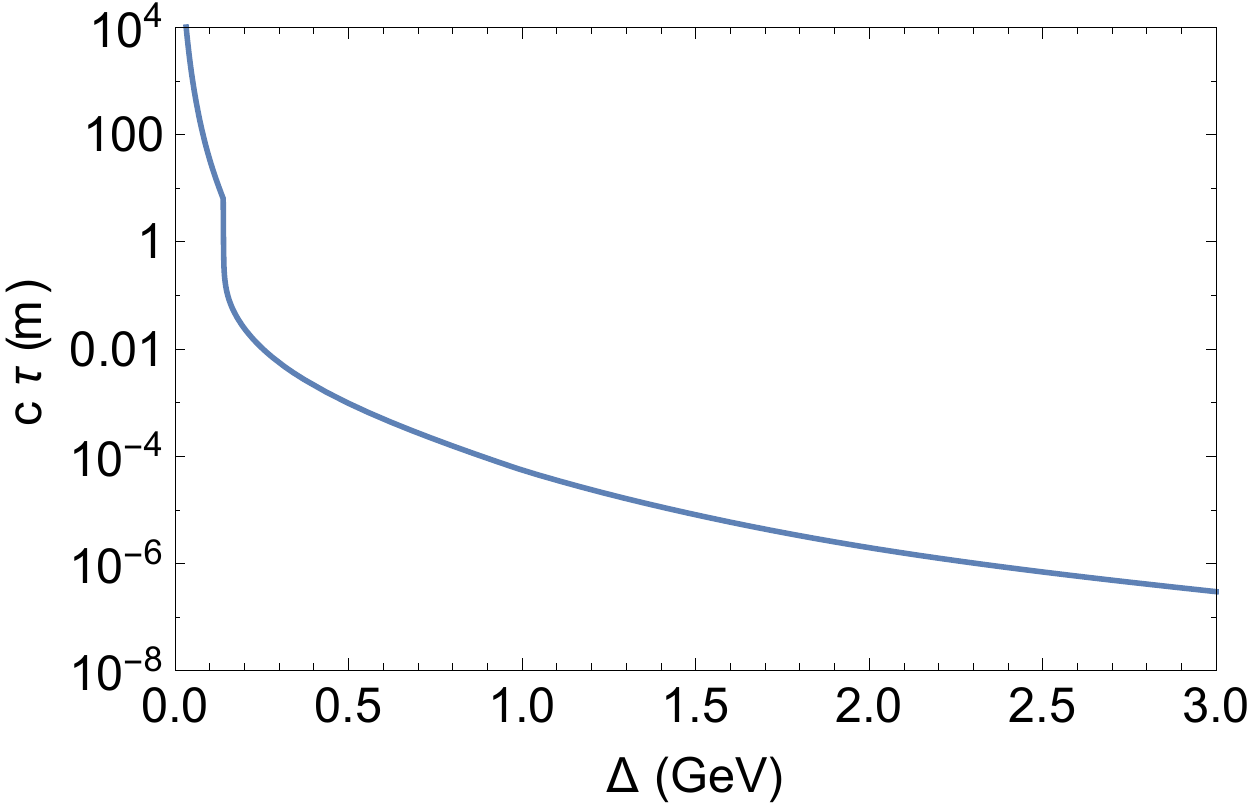}\quad
	\includegraphics[width=0.45\textwidth]{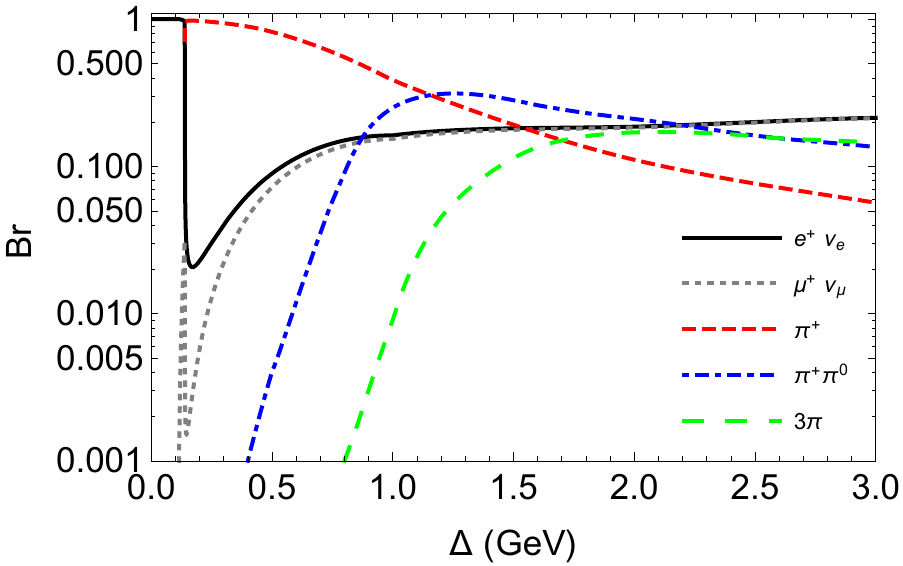}
	\caption{The lifetime (left) and branching ratios (right) for the triplet decay $\psi^+ \to \chi + SM$, as a function of the mass splitting $\Delta$.  As discussed in the text, all other decay widths are simple rescalings of this one, and the branching ratio is a universal function.}\label{fig:decay}
\end{figure}

The charged states of the multiplet will decay to the ground state through emission of virtual $W$s, $\psi^{q+} \to \psi^{(q-1)+} W^\ast$.  The lifetime is highly sensitive to the mass splitting, in particular because there are many thresholds for new decay modes in the GeV range.  For leptonic decay modes, we can easily compute the widths analytically; defining $\rho_l \equiv m_l/\Delta$,
\begin{equation}
	\Gamma (\psi^{q+} \to \psi^{(q-1)+} l^+ \nu_l) = \bigl( n^2 - (2q - 1)^2 \bigr) \, \frac{G_f^2 \Delta^5}{120 \pi^3} \, \biggl( \bigl( 2 - 9 \rho_{l}^2 - 8 \rho_{l}^4 \bigr) \sqrt{1 - \rho_{l}^2} + 15 \rho_{l}^4 \log \biggl[ \frac{1 + \sqrt{1 - \rho_{l}^2}}{\rho_{l}} \biggr] \biggr) \,,
\label{eq:pwlep}\end{equation}
where $\Delta = m_{\psi^{q+}} - m_{\psi^{(q-1)+}}$ and $n$ is the dimension of the multiplet.  The prefactor derives from the $W$ coupling to the DM multiplet.  As all decays proceed through this coupling, all partial widths and the total width have the same scaling with $n$ and $q$.  This allows us to compute the width as a function of $\Delta$ for one particle, and all remaining widths are given by an overall rescaling; while the branching ratio is a universal function of the mass splitting.  Additionally, the range between the smallest and largest widths we consider ($\psi^+ \to \chi W^\ast$ for the triplet and septet, respectively) is only a factor of 6.    \Figref{fig:decay} shows the lifetime and some of the largest branching ratios using the analytic expressions for the tau decay width used in \texttt{Herwig++}~\cite{0710.1951}.   As expected, the decay length covers many orders of magnitude for GeV-scale $\Delta$.  The most important features are the nearly-adjacent thresholds for decays to $\mu\nu_\mu$ and $\pi^+$ at $\Delta \gtrsim 100$\;MeV; when the relevant mass splitting is above this, the lifetime is at most a few cm, while below it we quickly have $c\tau > 100$\,m.

We have motivated our study of these models as candidates for thermal freeze out.  Because of their simplicity, the correct relic abundance is only obtainable for a narrow range of masses.  For the Higgsino-like doublet, we match observations for $m_\chi \approx 1$\;TeV.  Freeze-out of the other multiplets is sensitive to non-perturbative effects, including the Sommerfeld effect~\cite{Hisano:2006nn} and bound state formation~\cite{1407.7874}, resulting in the preferred mass depending on the mass splitting.  In all cases the relic density requires dark matter heavier than the kinematic limit of 1.5\;TeV.  However, there is still value in considering lighter masses; freeze-out in this case \emph{under-produces} DM, such that they are not excluded and could be part of a multi-component DM theory.  For all these reasons we will not impose any relic density constraint.

%% file: Files/evgen.tex
In the following sections, we will derive the discovery potential and prospective constraints on the models outlined above from direct searches at a future $e^+e^-$ collider.  As discussed in the previous section, the lifetime of the charged states that are produced can vary over many orders of magnitude for mass splittings in the GeV-range.  A number of different channels must be considered in order to effectively cover the mass-lifetime plane.  The particular details of the different search strategies are given below; but first we outline some technical details that are common to all.

The need to pair-produce dark sector particles means that the absolute maximum mass (the \emph{kinematic limit}) that can be probed by direct searches is $m_\chi = \sqrt{s}/2$.  The current LHC constraints on a pure Wino are already $m_\chi \gtrsim 460$\,GeV~\cite{1712.02118}, demanding that we consider colliders with $\sqrt{s} > 1$\,TeV.  CLIC has the highest centre of mass energy among current proposals for linear colliders, and so we focus on this experiment.  Except where noted, the specifications of the accelerator and detector are taken from the CLIC conceptual design report (CDR)~\cite{1202.5940}.  This includes the centre of mass energy $\sqrt{s} = 3$\,TeV, the lifetime integrated luminosity of 2\,ab$^{-1}$, and the beam polarisation of 80\% (30\%) for $e^-$ ($e^+$).  Following \refcite{1506.07830}, we assume that when operating in discovery mode, the integrated luminosity is split over the four different helicity combinations as shown in \tabref{tab:lumi}.  When setting limits, we compare those found using all data with those using only specific initial polarisations, and choose the strongest.

\begin{table}
	\centering
	\begin{tabular}{|c|c|}
		\hline
		$P(e^-)$, $P(e^+)$ & Luminosity fraction \\
		\hline
		(+, +) & 10\% \\
		(+, $-$) & 40\% \\
		($-$, +) & 40\% \\
		($-$, $-$) & 10\% \\
		\hline
	\end{tabular}
	\caption{Luminosity fraction for the four beam polarisation configurations, taken from~\refcite{1506.07830}.}\label{tab:lumi}
\end{table}

We generate both signal and background events with these parameters using Whizard~2.4.0~\cite{0708.4233,hep-ph/0102195,1010.3251}, with the contributions of bremsstrahlung and beamsstrahlung computed using GuineaPig++~\cite{Schulte:2007zz}.  For the signal events we first implement our models in FeynRules~2.3.26~\cite{1310.1921,0806.4194}.  Beamsstrahlung and bremsstrahlung have a significant effect for the CLIC beam parameters; only 30\% of collisions occurring at the nominal centre of mass energy, and the incident beams contain a large fraction of energetic photons.   For this reason, in addition to $e^+e^-$-initiated events, we include contributions from $e^\pm\gamma$ and $\gamma\gamma$ initial states for both signal and background.

The precise details of the detector are highly likely to change before construction and operation.  We therefore do not attempt a full detector simulation.  Rather, we consider two possibilities: a best-case outcome based on truth-level Monte Carlo output; and a simple estimate of the impact of reconstruction efficiencies and energy smearing.  Even in the ideal case, we impose two cuts on all final state objects based on expected features of the detector.  The first is an angular cut $\lvert\cos\theta\rvert < 0.99$ ($\lvert \eta \rvert < 2.65$) that corresponds to the planned physical dimensions of the machine.  The second is a cut on the transverse momemtum $p_T > 10$\;GeV, so as to avoid contamination from the large $\gamma\gamma \to$\,hadrons pile up expected at CLIC  (an average of 19\,TeV per bunch crossing, of which 1.2\,TeV is coincident with a single event readout).  These cuts also apply to the more realistic detector modelling; we augment them with energy-dependent reconstruction efficiencies that average 93\% for photons, 97\% for electrons and 99\% for muons.  Objects that pass our cuts and are not reconstructed are assumed to leave no detector signal (no fake rate is applied).  The energies are then smeared by a Gaussians with energy-dependent widths.  For photons, the width is simply given by
%
\begin{equation}
	\frac{\sigma^\gamma_E}{E} = 1.089\% \oplus \frac{16.69\%}{\sqrt{E/\text{GeV}}} \,,
\end{equation}
with the two components of the uncertainty added in quadrature.  For charged leptons, the energy resolution is best fit using a sum of two Gaussians, with widths
\begin{align}
	\frac{\sigma^{e,1}}{E^2} & =1.4 \times 10^{-5} \, \text{GeV}^{-1} \,, & \frac{\sigma^{\mu,1}}{E^2} & =1.5 \times 10^{-5} \, \text{GeV}^{-1} \,, \\
	\frac{\sigma^{e,2}}{E^2} & =7.7 \times 10^{-5} \, \text{GeV}^{-1} \,, & \frac{\sigma^{\mu,2}}{E^2} & =4.9 \times 10^{-5} \, \text{GeV}^{-1} \,.
\end{align}
For electrons (muons), the narrower Gaussian has weight 70\% (95.9\%).  Finally, for final states that involve more than one hard particle we impose a separation cut $\Delta R > 0.4$.

The majority of our searches have a non-zero background.  In this case, our discovery and exclusion criteria are based on a simple significance function.  Given $N_{sig}$, $N_{bkg}$ expected signal and background events, the significance $\mathcal{S}$ is given by
\begin{equation}
	\mathcal{S} = \frac{N_{sig}}{\sqrt{N_{bkg} + (\epsilon_{sys} \, N_{bkg})^2}} \,.
\label{eq:sig}\end{equation}
This is the ratio between the number of signal events and the uncertainty on the background, where the latter is given by the sum in quadrature of the statistical uncertainty $\sqrt{N_{bkg}}$ and the systematic uncertainty $\epsilon_{sys} \, N_{bkg}$.  Our choice for the size of the systematic factor $\epsilon_{sys}$ will depend on the search and is discussed in more detail below.  Our discovery (exclusion) criteria is $\mathcal{S} = 5$ (2).  For the search in \secref{sec:dtrack} where we expect a very small and possibly zero background, we adopt conservative criteria of 10 events for discovery and 5 expected events for exclusion.

%% file: Files/Muons.tex
When the charged states have a sufficiently large lifetime, they can survive long enough to exit the experiment.  They leave a signal in the muon chambers, but can be distinguished from true muons by their velocity $\beta$ inferred from either the time of flight or the radius of curvature in the solenoid magnetic field.  The CLIC detectors are expected to have a radius of $\sim$10\,m transverse to the beam axis, which sets an approximate lower bound on $c\tau$ for this search to be effective.  As can be seen from \figref{fig:decay}, this corresponds to mass splittings below the muon threshold, $\Delta \lesssim 100$\;MeV.  For the larger multiplets, in the mass range 30\;MeV\;$\lesssim \Delta m_1 \lesssim 100$\;MeV only the singly-charged state is long-lived, with the other unstable particles decaying promptly.  We will first assume this to be the case, then discuss how the presence of multiple long-lived states might modify our constraints in \secref{sec:comb}.

Since all the mass splittings are sub-GeV, the SM decay products are soft and difficult to reconstruct from the coincident $\gamma\gamma \to$~hadrons objects.  While we could use their presence as an additional handle to distinguish signal from background for the larger multiplets, to be conservative we will assume these decay products can not be resolved, and base our search purely on the existence of long-lived charged particles (LLCPs).  Searches for LLCPs have been performed at ATLAS~\cite{1411.6795,1808.04095} and CMS~\cite{1305.0491,1609.08382}, and we use their analyses as a guide.  In particular, we focus on a signal of two hard LLCP tracks satisfying the cuts of \refcite{1411.6795}:
\begin{itemize}
	\item $p_T > 70$\;GeV;
	\item $\lvert\eta\rvert < 2.5$;
	\item $\beta < 0.95$;
\end{itemize}
and our usual isolation cut $\Delta R > 0.4$.  We weight events by the probability that both LLCPs travel a transverse distance of at least $L_0 \equiv 20$\,m prior to decaying,
\begin{equation}
	w^{dec} = \prod_i e^{-L_0/\gamma_i c \tau \cos\theta_i} \,,
\label{eq:LLCPwgt}\end{equation}
where $\gamma_i$ and $\theta_i$ are the boosts and polar angles, respectively, of the two LLCPs.  The signal production rate is dominated by the simple $s$-channel $e^+e^- \to \gamma^{\ast}/Z^\ast \to \psi^{q+}\psi^{q-}$ shown in \figref{fig:prod}.  Since the $\gamma/Z$ coupling to the dark sector is proportional to the charge, the total production cross sections scale as the sum of the squares of the charges of the states in the multiplet; the ratio triplet\,:\,quintet\,:\,septet is $1^2: 1^2 + 2^2: 1^2 + 2^2 + 3^2 = 1:5:14$.  (The Higgsino does not follow this pattern, as it has non-zero hypercharge, but we expect its production cross section to be the smallest among fermions.)  We also include the contributions from $\gamma\gamma$-initiated events and, for the larger multiplets, charge-asymmetric channels such as $e^\pm\gamma \to \nu W^{\ast} \to \nu \psi^{q\pm} \psi^{(q-1)\mp}$.  These are small and so do not significantly modify the production ratio.  When we include detector effects, we use muon detection as a proxy for the LLCP reconstruction efficiency.  We also considered an alternative search strategy with one LLCP and a hard photon; however, since LLCPs are almost always pair produced\footnote{The exception is $e^\pm \gamma \to \psi^{\pm} \chi \nu$, which is never the dominant production channel.}, this is inferior.

The dominant background to this search comes from muons with mis-measured velocities, with other sources negligible.  This is a difficult background to estimate, since it depends on the precise details of the detector performance.  We make the conservative estimate that the CLIC detectors will be able to do at least as well as those at the LHC, and use an estimated fake rate from \refcite{1411.6795}.  The expected background was below 1 event for $m_\chi > 200$\;GeV for 19.1\;fb$^{-1}$ of data, compared to a muon pair-production cross section after the $p_T$ and angular cuts of 0.76\;pb.  This corresponds to a very conservative estimated fake rate
\begin{equation}
	\mathcal{P}^{fake} < \frac{1}{19.1 \times 0.76 \times 10^3} = 7 \times 10^{-5}\,.
\label{eq:LLCPfake}\end{equation}
This implies a background cross section of $\approx 0.1$\;fb.  Additionally, as the background derives from vector-like QED processes it varies with the beam polarisations only very weakly.

\begin{figure}
	\centering
	\includegraphics[width=0.45\textwidth]{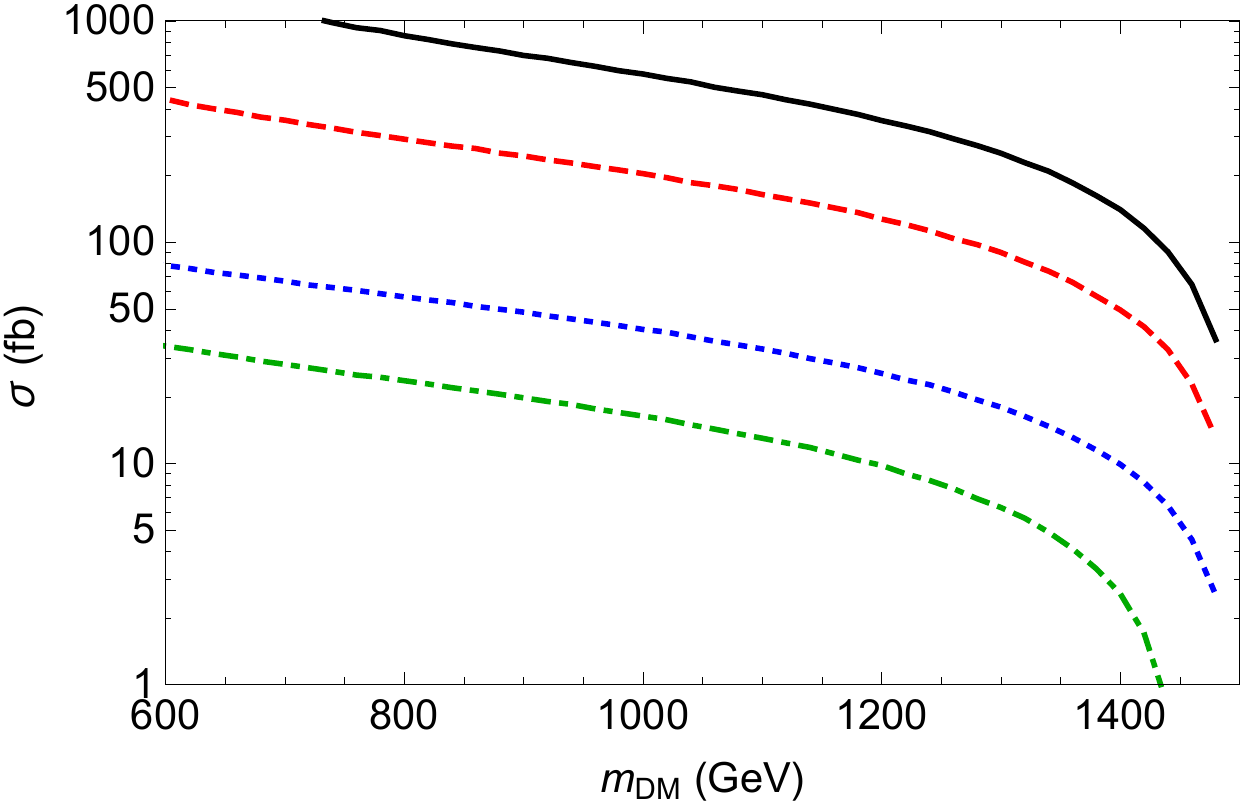}\quad
	\includegraphics[width=0.45\textwidth]{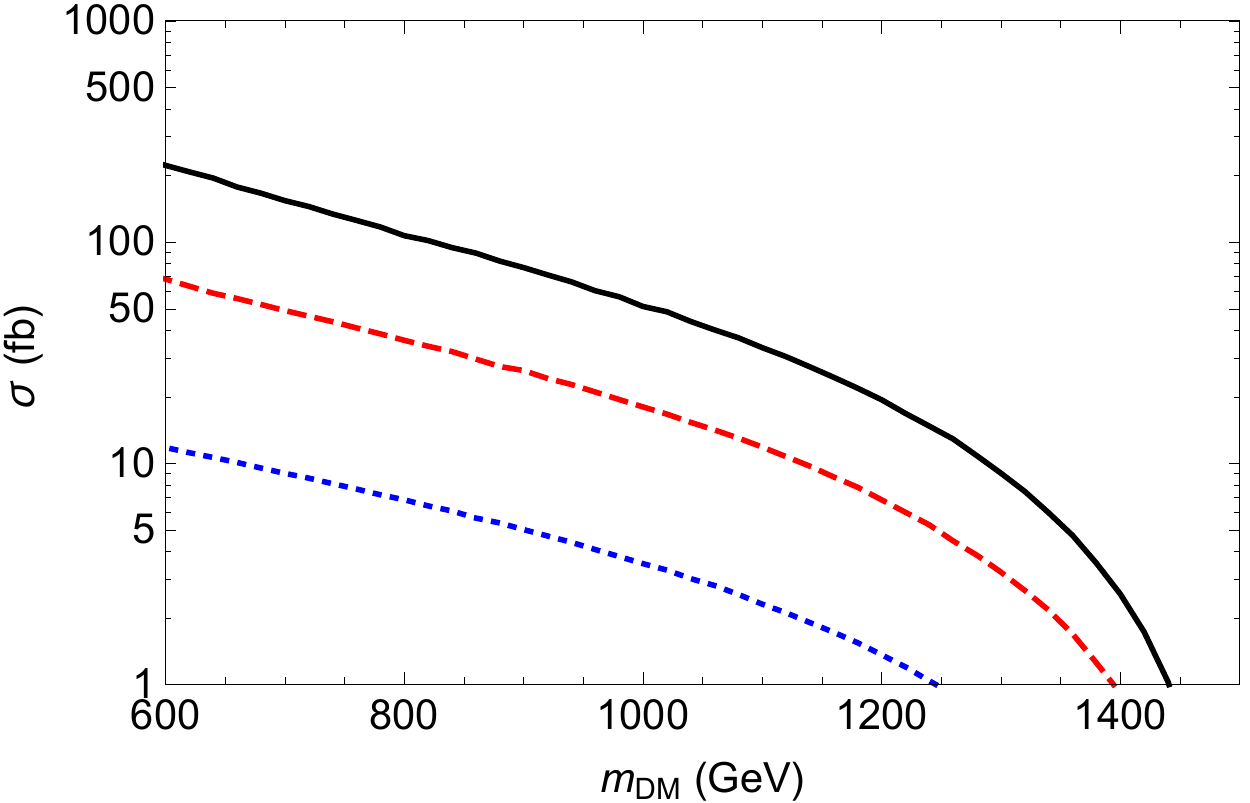}
	\caption{Cross sections (after cuts and detector effects) for the LLCP signals, for fermions (left) and scalars (right).  The black solid (red dashed, blue dotted, green dot-dashed) line shows the result for the septet (quintet, triplet, doublet) representation.  For comparison, the background cross section is 0.15\;fb.}\label{fig:LLCPxsec}
\end{figure}

\begin{figure}
	\centering
	\includegraphics[width=0.45\textwidth]{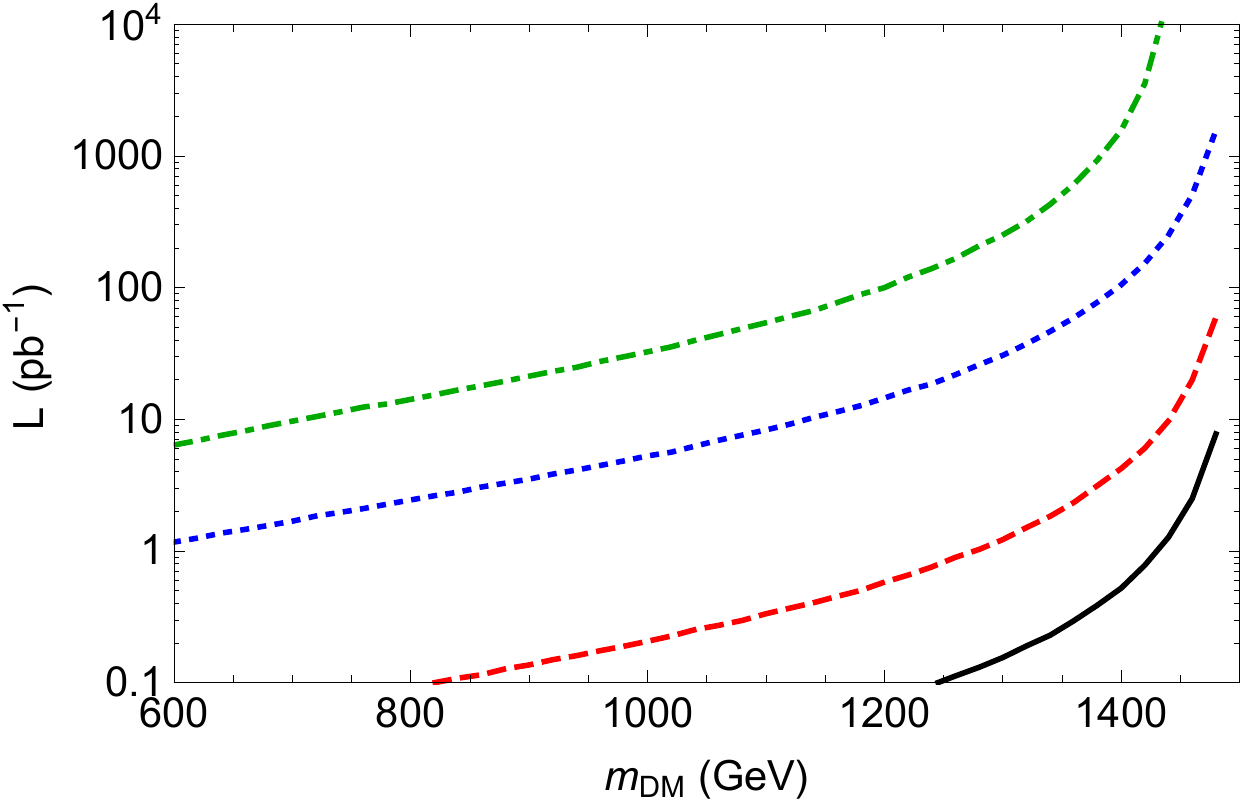}\quad
	\includegraphics[width=0.45\textwidth]{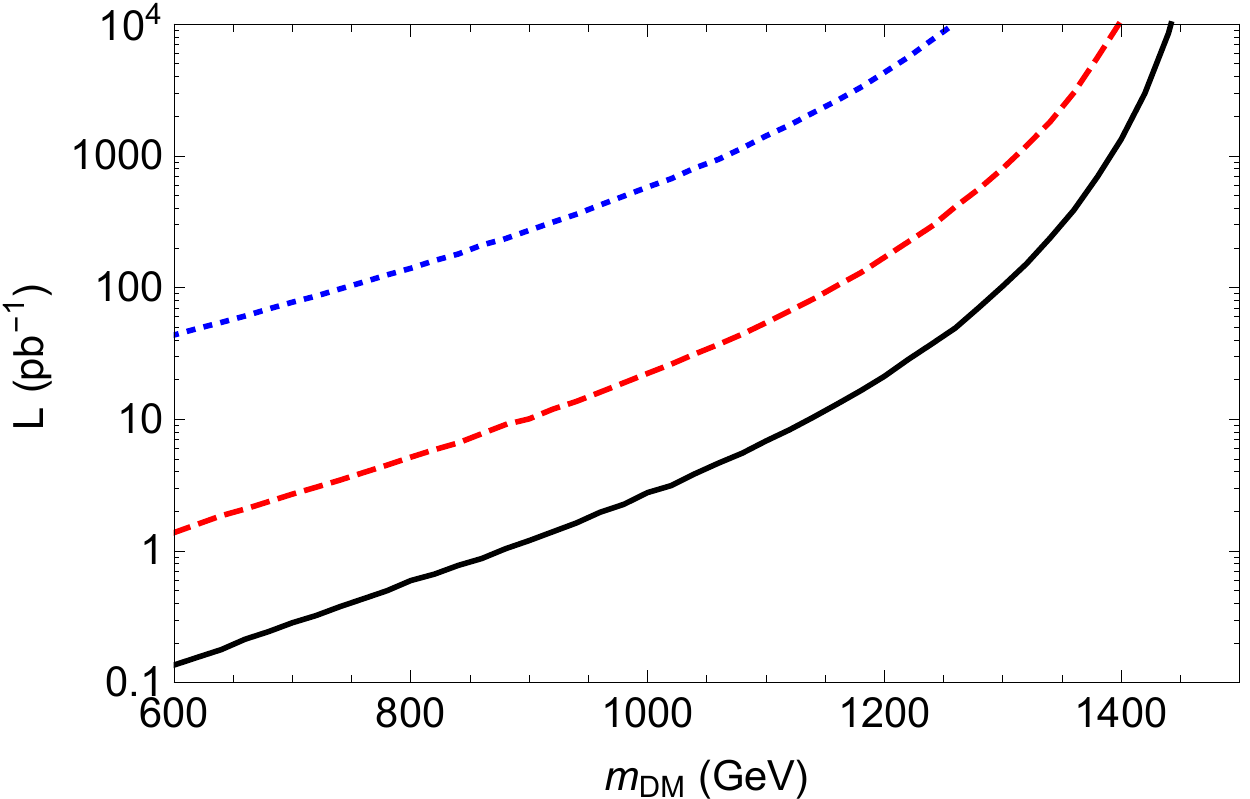}
	\caption{Luminosity required for a 5$\sigma$ discovery for LLCPs with lifetime $c\tau = 10$\;m for fermions (left) and scalars (right).  The different lines have the same representation as in \figref{fig:LLCPxsec}.  We see that for all fermion models but the doublet, as well as for the larger scalar multiplets, we can approach the kinematic limit 1.5\;TeV with only a fraction of the design luminosity.}\label{fig:LLCPlum5}
\end{figure}

In contrast, the signal cross section has a strong dependence on polarisation since it proceeds through the $SU(2)_L$ weak coupling.  The best sensitivity arises when we exploit this by only considering the polarisation $\bigl(P(e^-)$, $P(e^+)\bigr) = (-, +)$.  The signal cross section is maximised for this choice, at least twice the rate for each other beam polaristaion.  We show these cross sections after applying cuts and detector effects as a function of mass in \figref{fig:LLCPxsec}; they approximately obey the $1:5:14$ ratio discussed above.  The fermion cross section is enhanced over the scalar one by a factor of 2 for degrees of freedom; and additionally by the need to produce the scalars in a $L=1$ state.  In the centre of mass frame\footnote{Recall that due to beamsstrahlung, this frame will be boosted along the beam axis for a large fraction of events.} the tree-level cross sections for $e^+e^- \to \psi^{q+}\psi^{q-}$ are
\begin{align}
	\frac{d\sigma}{d\cos\theta} \biggr\rvert_{fer} & = \frac{\pi \alpha^2 q^2}{4E^2} \, \frac{p}{E} \, \biggl(2 - \frac{p^2}{E^2} \sin^2\theta \biggr) \,, \\
	\frac{d\sigma}{d\cos\theta} \biggr\rvert_{sca} & = \frac{\pi \alpha^2 q^2}{8E^2} \, \frac{p^3}{E^3} \sin^2\theta \,,
\end{align}
where $\theta$ is the polar angle of the final state, $E$ the energy and $p$ the three-momentum.  We can see that the lack of an $s$-wave final state reduces the scalar cross section, especially at high masses since the matrix element is proportional to the velocity $p/E$.  The high muon tagging efficiency combined with the hard $p_T$ cuts means that there is very little difference in the truth-level cross sections, so we do not show them.  It is clear that except near the kinematic limit, the signal cross section is much larger than the background allowing strong limits to be set even in the presence of a large systematic uncertainty.  In \figref{fig:LLCPlum5} we show the integrated luminosity required for a 5$\sigma$ discovery assuming $\epsilon_{sys} = 50\%$, for a singly-charged state decay length of 10\;m.  Except for the smallest multiplets, we reach $m_\chi = \sqrt{s}/2$ in only a fraction of the design luminosity of 2\;ab$^{-1}$.  The exclusion contours are obviously even stronger.  For the expected signal reach in the mass-lifetime plane, see our composite plots in \secref{sec:comb}.

%

%% file: Files/Tracks.tex
When the charged states in the EWDM multiplet have a lifetime $c\tau \sim 1$\;cm---10\,m, they survive long enough to enter the detector volume, but decay before reaching the muon chambers.  They will leave tracks within the tracking system as they carry electric charge, but their large mass means they deposit little to no energy in the calorimeters.  The result is a ``disappearing tracks'' signal with a background from detector fakes only, \emph{i.e.} essentially no SM background.  This is particularly relevant as lifetimes of this order are expected when the mass splitting in the multiplet derives purely from radiative corrections.  We can see from \figref{fig:decay} and~\modeqref{eq:dmhighq} that when the singly-charged state has a lifetime in this range, any other unstable states will decay promptly; while if the higher-charged states live this long, the singly-charged state will be collider-stable.  We focus on the former case, and discuss the possible effects of the latter in \secref{sec:comb}.

Current searches at the LHC cannot directly trigger on the disappearing track signal; additional hard objects are required~\cite{ATLAS:2017bna}.  The presence of these energetic states, and the consequent large missing momentum, also help to suppress the fake rate.  In contrast, it is proposed to record all data at CLIC~\cite{1202.5940}, so no trigger is required.  Provided that the background from fakes is well-understood, it might be possible to set limits using \emph{only} events with one or two disappearing tracks and no other hard objects.  However, since we cannot at this stage know what that background is, we conservatively use LHC-like events to set limits.

\begin{figure}
	\centering
	\begin{tikzpicture}[node distance=0.4cm and 0.4cm]
		\coordinate (v1);
		\coordinate[right = of v1] (va);
		\coordinate[right = of va, label=below:$Z/\gamma$] (v2);
		\coordinate[right = of v2] (vb);
		\coordinate[right = of vb] (v3);
		\coordinate[above left = of v1] (vi1);
		\coordinate[above left = of vi1, label=above left:$e^-$] (i1);
		\coordinate[below left = of v1] (vi2);
		\coordinate[below left = of vi2, label=below left:$e^+$] (i2);
		\coordinate[above right = of v3] (vo1);
		\coordinate[above right = of vo1, label=above right:{$\psi^{q-}$}] (o1);
		\coordinate[below right = of v3] (vo2);
		\coordinate[below right = of vo2, label=below right:{$\psi^{q+}$}] (o2);
		\coordinate[left = of o1] (oa);
		\coordinate[left = of oa, label=above right:{$\gamma$}] (o3);
		\draw[fermion] (i1) -- (vi1);
		\draw[fermion] (vi1) -- (v1);
		\draw[fermion] (v1) -- (i2);
		\draw[photon] (v1) -- (v3);
		\draw[fermionnoarrow] (o2) -- (v3);
		\draw[fermionnoarrow] (v3) -- (o1);
		\draw[photon] (vi1) -- (o3);
	\end{tikzpicture}\qquad\qquad
	\begin{tikzpicture}[node distance=0.4cm and 0.4cm]
		\coordinate (v1);
		\coordinate[right = of v1] (va);
		\coordinate[right = of va, label=below:$Z/\gamma$] (v2);
		\coordinate[right = of v2] (vb);
		\coordinate[right = of vb] (v3);
		\coordinate[above left = of v1] (vi1);
		\coordinate[above left = of vi1, label=above left:$e^-$] (i1);
		\coordinate[below left = of v1] (vi2);
		\coordinate[below left = of vi2, label=below left:$e^+$] (i2);
		\coordinate[above right = of v3] (vo1);
		\coordinate[above right = of vo1, label=above right:{$\psi^{q-}$}] (o1);
		\coordinate[below right = of v3] (vo2);
		\coordinate[below right = of vo2, label=below right:{$\psi^{q+}$}] (o2);
		\coordinate[right = of vo1] (oa);
		\coordinate[right = of oa, label=right:{$\gamma$}] (o3);
		\draw[fermion] (i1) -- (v1);
		\draw[fermion] (v1) -- (i2);
		\draw[photon] (v1) -- (v3);
		\draw[fermionnoarrow] (o2) -- (v3);
		\draw[fermionnoarrow] (v3) -- (o1);
		\draw[photon] (vo1) -- (o3);
	\end{tikzpicture}
	\caption{Contributing Feynman diagrams for $e^+e^- \to \psi^{q+} \psi^{q-} \gamma$.  (Left): initial state radiation.  (Right): final state radiation.}\label{fig:isrfsr}
\end{figure}
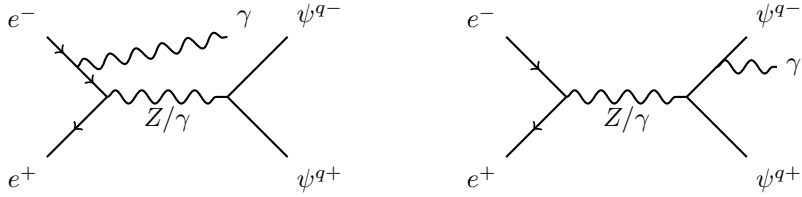

We therefore consider a final state of two members of the DM multiplet, plus one visible sector state.  We expect the largest rate for processes with the $e^+e^-$-initial state, so the visible final state particle must be a boson.  The photon, being massless and coupling to both the initial and final states, will dominate.  The relevant processes are shown in \figref{fig:isrfsr}.  Usually we expect initial state radiation (ISR) to be more important due to the collinear and soft singularities.  However, these effects are greatest in regions of parameter space that we remove with angular and $p_T$ cuts.  Final state radiation (FSR) is enhanced by an additional factor of $q^2$ over ISR, making its contribution relatively more significant for the larger multiplets.  Especially for lighter DM, these two effects might lead to the overall cross section scaling as $q^4$, with the subsequent ratio triplet\,:\,quintet\,:\,septet of $1^4:1^4 + 2^4:1^4 + 2^4 + 3^4 = 1:17:98$.  Realistically, due to the combination of both ISR and FSR we expect the scaling to lie somewhere between this value and the $1:5:14$ ratio of the previous section.

We impose the following signal cut:
\begin{itemize}
	\item At least one photon with $p_T > 100$\; GeV;
	\item At least one charged state $\psi^\pm$ in the detector volume with $p_T > 25$\;GeV.
\end{itemize}
The latter is our candidate disappearing track.  For this to register as our signal, it must live long enough to leave a reconstructable track but decay before entering the muon chambers, to avoid misidentification as a muon.  The precise distances these correspond to will again depend on the details of the detector design.  For the former, we follow LHC searches~\cite{ATLAS:2017bna} and demand that $\psi^\pm$ travel at least 10\;cm transverse to the beam axis; while for the latter, we assume the muon chambers begin at a radial distance of 4\;m~\cite{1202.5940}.  The probability of $\psi^\pm$ producing a disappearing track is therefore given by
\begin{equation}
	\mathcal{P}^{tr} = \mathcal{P}^{dec} (\gamma, \theta) \times \mathcal{P}^{rec} \,,
\end{equation}
where $\mathcal{P}^{dec}$ is the decay probability as a function of the boost $\gamma$ and polar angle $\theta$,
\begin{equation}
	\mathcal{P}^{dec} (\gamma, \theta) = \exp \biggl( - \frac{\text{10 cm}}{\gamma c\tau \tan\theta} \biggr) - \exp \biggl( - \frac{\text{4 m}}{\gamma c\tau \tan\theta}\biggr) \,,
\end{equation}
and $\mathcal{P}^{rec}$ is the reconstruction efficiency for the disappearing track.  We consider values of this parameter between an optimistic 100\% and a pessimistic 10\% rate.  This event weight is further multiplied by the photon reconstruction efficiency.

\begin{figure}
	\centering
	\includegraphics[width=0.41\textwidth]{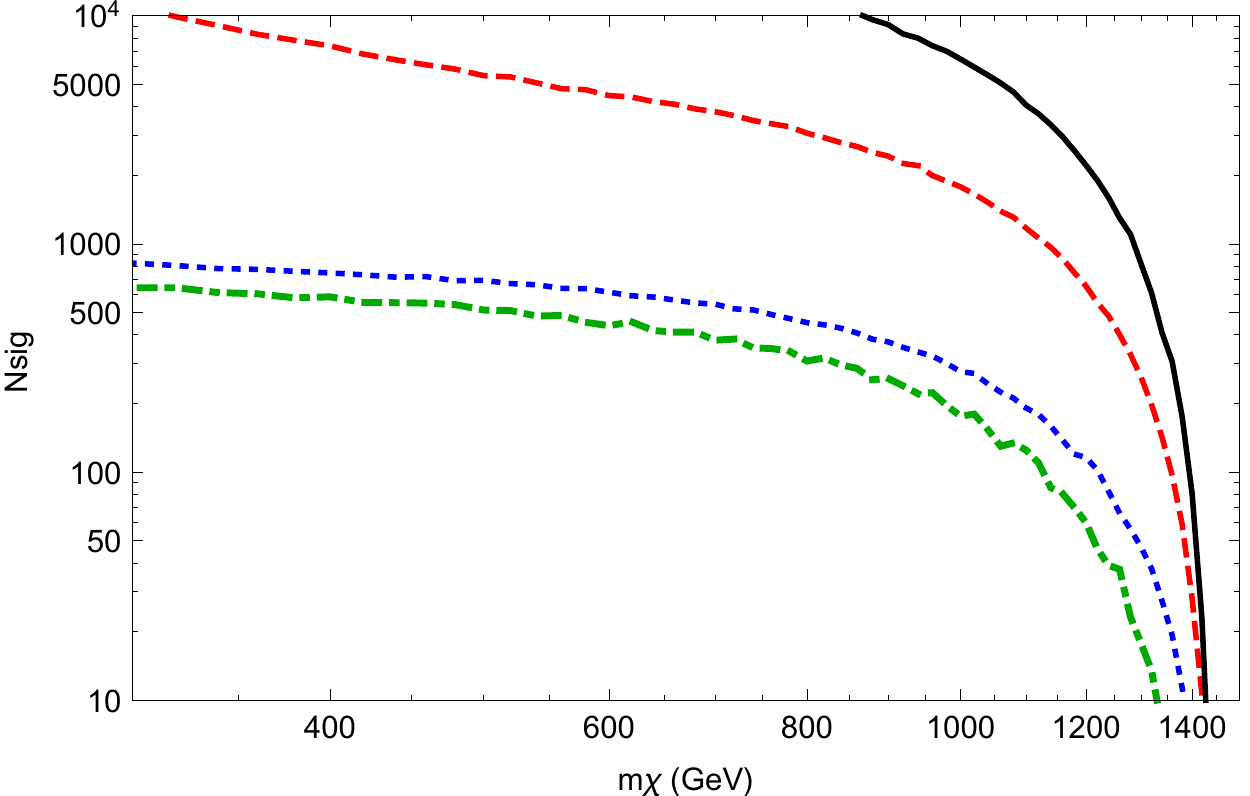}\quad
	\includegraphics[width=0.41\textwidth]{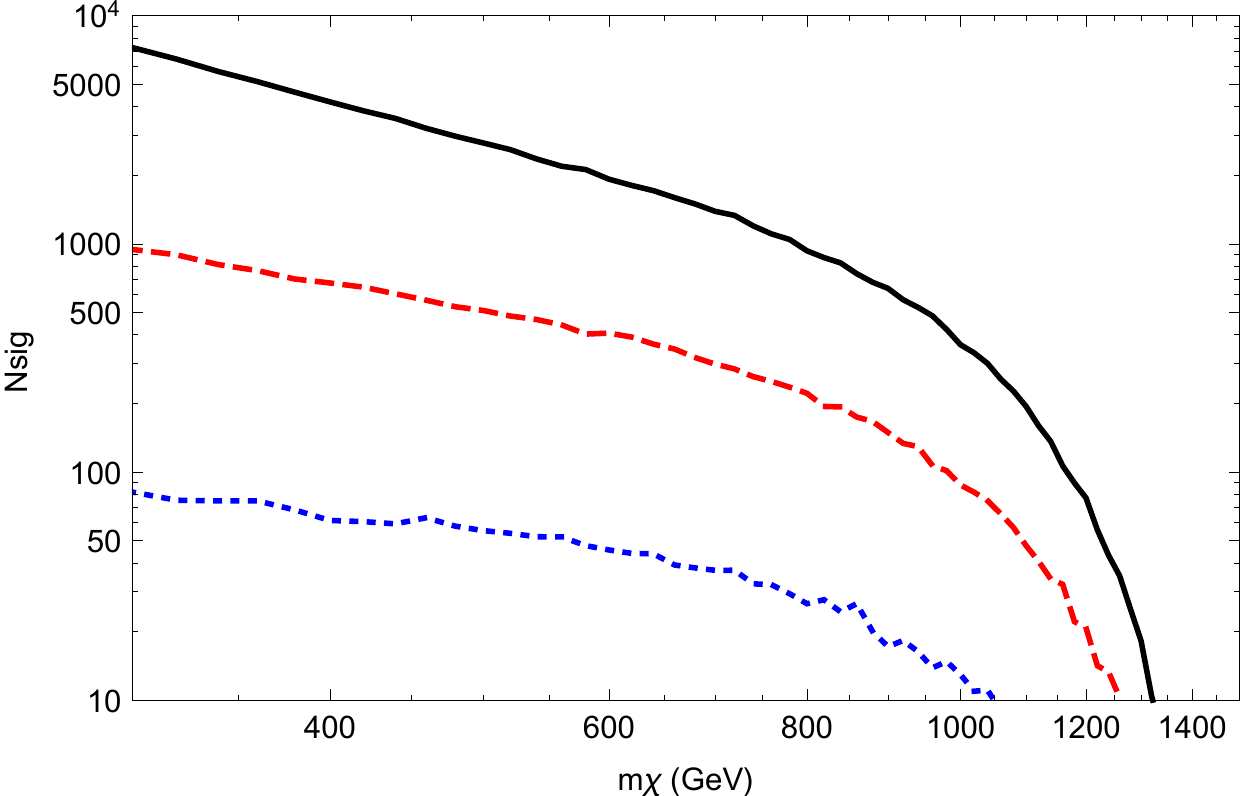}\\
	\includegraphics[width=0.41\textwidth]{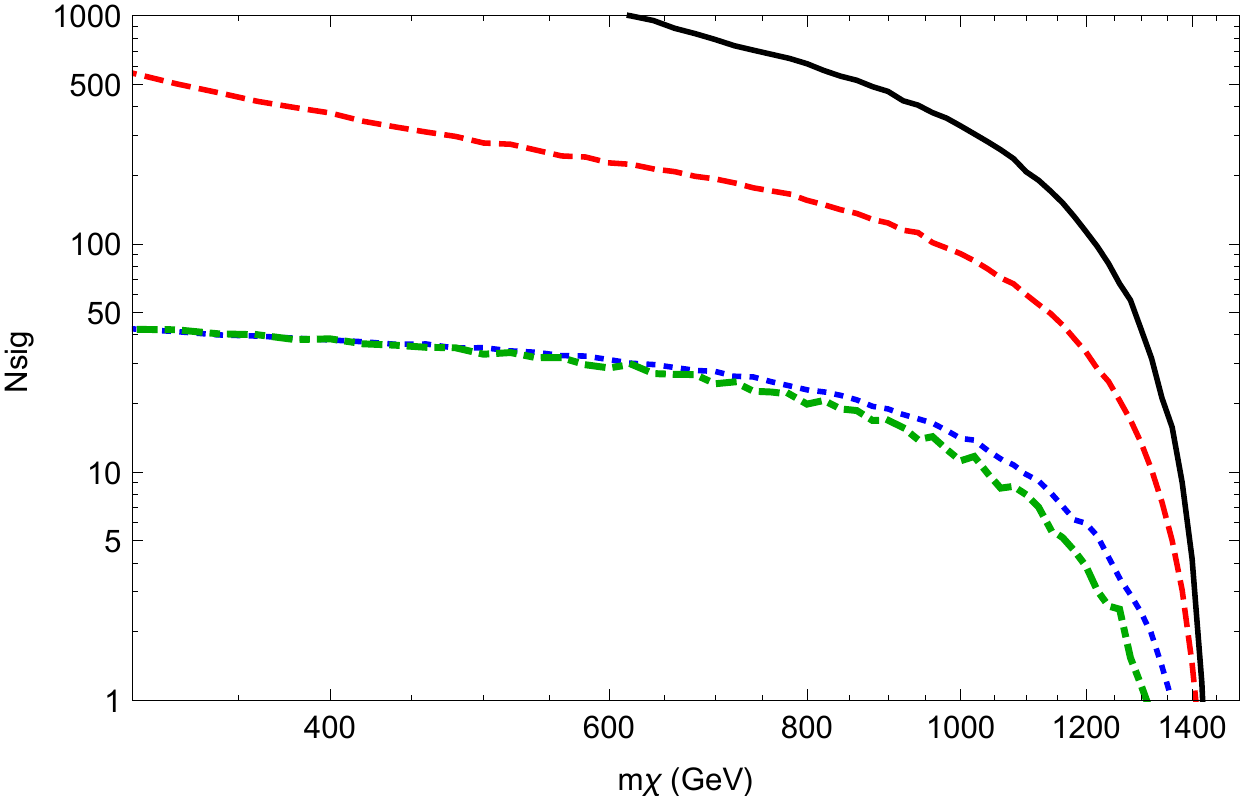}\quad
	\includegraphics[width=0.41\textwidth]{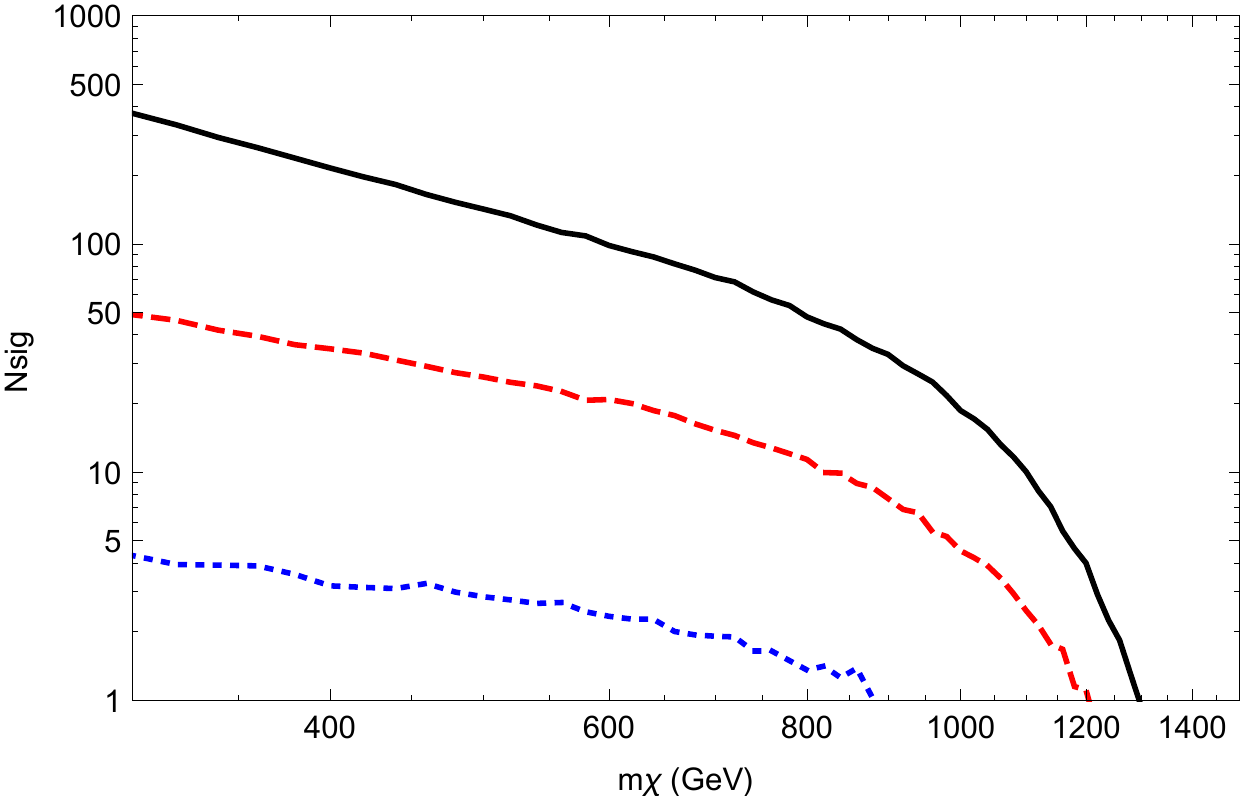}\\
	\caption{Number of single-disappearing-track events expected in our models as a function of mass.  The left (right) side shows fermion (scalar) models.  The upper row assumes 100\% efficiency for photon and disappearing track reconstruction; the lower row assumes the energy-dependent photon efficiency and 10\% for disappearing tracks.  The different lines label models with the same convention as \figref{fig:LLCPxsec}.}\label{fig:dtr_sing}
\end{figure}

\begin{figure}
	\centering
	\includegraphics[width=0.41\textwidth]{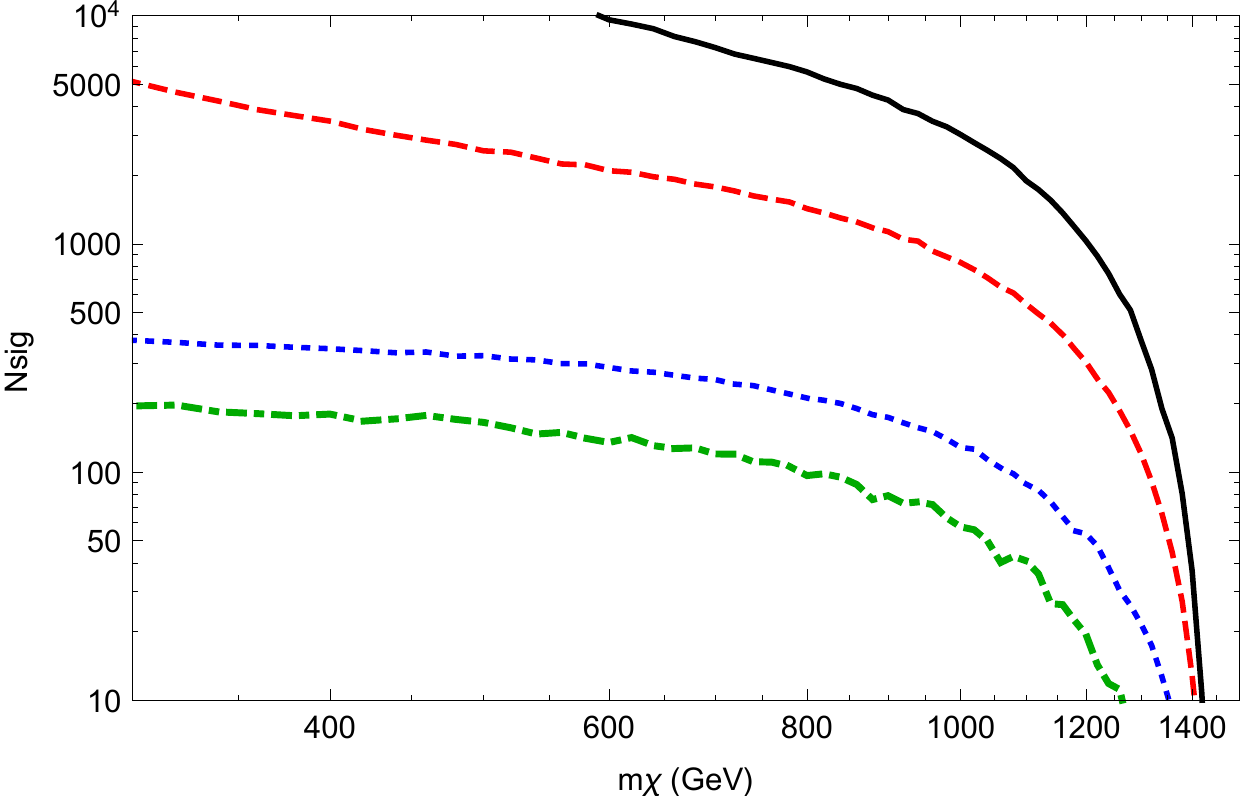}\quad
	\includegraphics[width=0.41\textwidth]{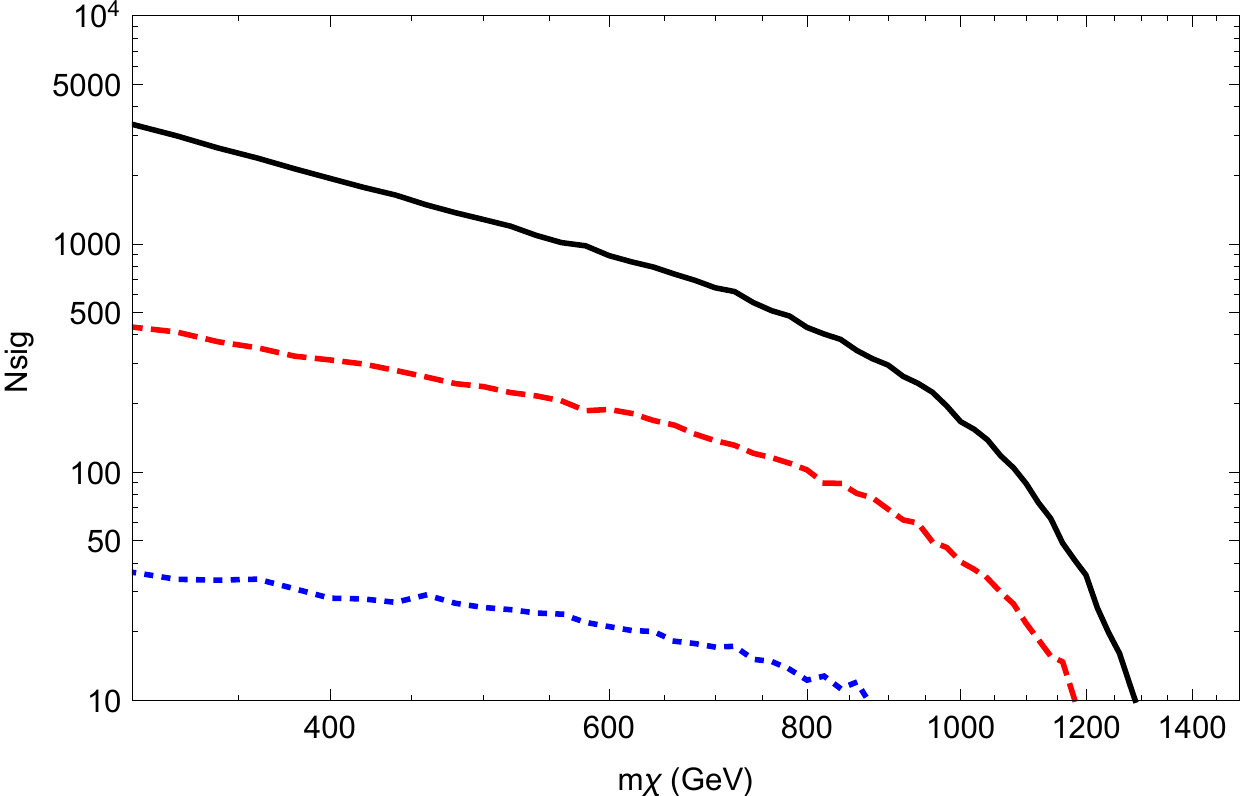}\\
	\includegraphics[width=0.41\textwidth]{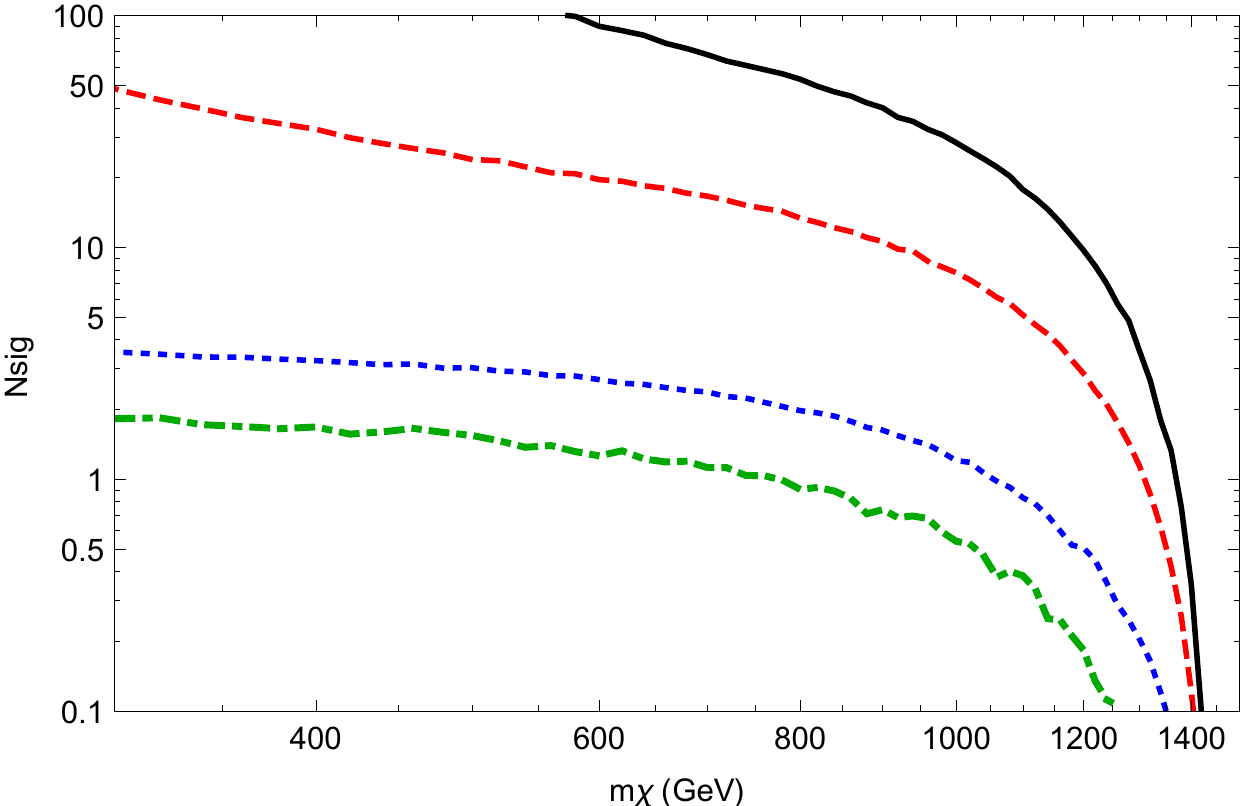}\quad
	\includegraphics[width=0.41\textwidth]{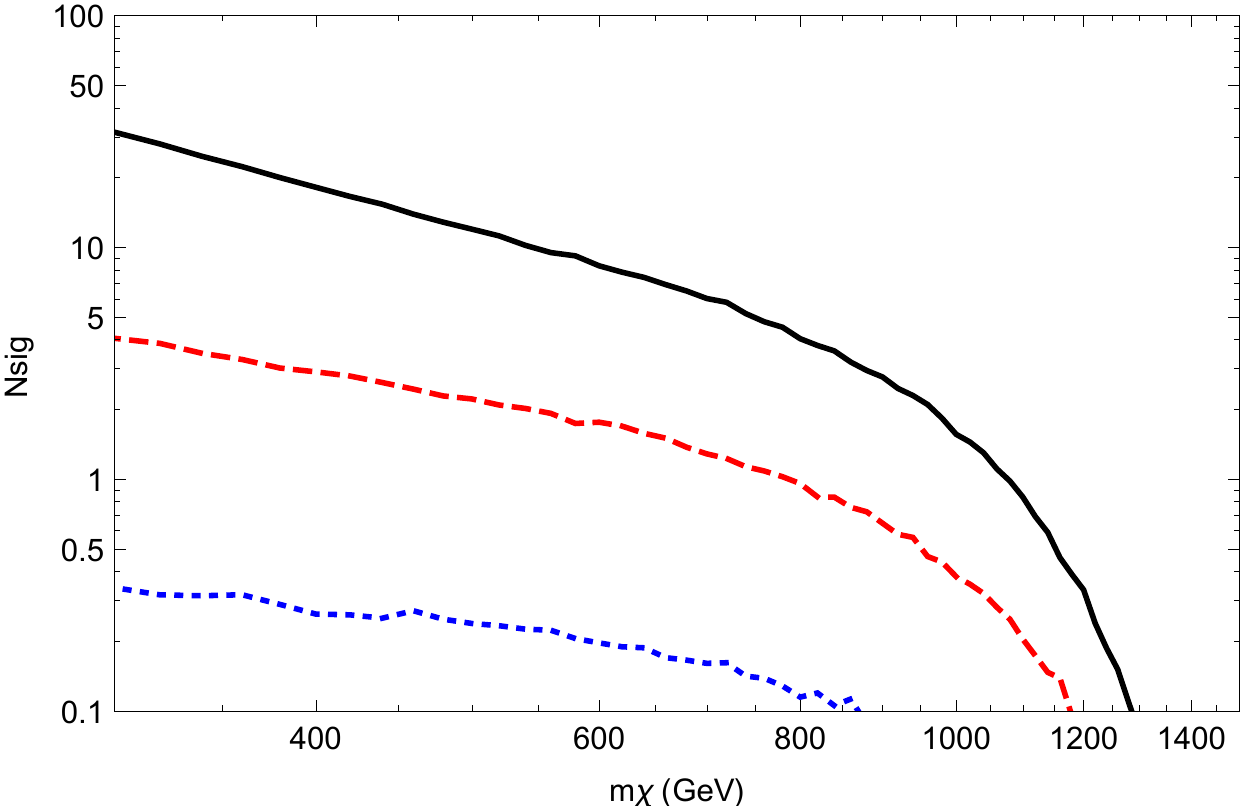}\\
	\caption{Number of double-disappearing-track events expected in our models as a function of mass.  The left (right) side shows fermion (scalar) models.  The upper row assumes 100\% efficiency for photon and disappearing track reconstruction; the lower row assumes the energy-dependent photon efficiency and 10\% for disappearing tracks.  The different lines label models with the same convention as \figref{fig:LLCPxsec}.}\label{fig:dtr_dbl}
\end{figure}

As noted above, the background to this signal depends upon details of the detector performance that we can not reliably predict.  Our choice of the hard photon cut is to suppress fakes by making it exceedingly unlikely that any SM particle could leave a track without also depositing substantial calorimeter or muon chamber activity.  This is especially likely to be the case if we demand the presence of \emph{two} disappearing tracks; since our signal is dominated by $e^+e^- \to \psi^{q+}\psi^{q-}\gamma$, the main suppression this has on our signal is an additional factor of $\mathcal{P}^{rec}$.  We therefore assume a zero-background search with a discovery (exclusion) criteria of 10 (5) events, where we sum over all beam polarisations.  In \figrefs{fig:dtr_sing} and~\ref{fig:dtr_dbl}, we show the number of expected events as a function of mass for our models for single- and double-disappearing track signals for a $\psi^+$ lifetime $c\tau = 1$\;m.  We see that the cross sections grow much more rapidly with increasing multiplet size than in the previous section.  Our hard photon cut is making FSR relatively more important, and the predicted ratio triplet\,:\,quintet\,:\,septet of $1:17:98$ is close to what is observed in \figrefs{fig:dtr_sing} and~\ref{fig:dtr_dbl}.  The ratios of the scalar to fermion cross sections are again much smaller than the factor of two expected from the degrees of freedom due to the different kinematic structure of the photon-multiplet couplings.  We see that if a single-track signal is sufficient, then even in our most pessimistic scenario we have a discovery potential of at least 1\;TeV for all models except the scalar triplet.  However, if two disappearing tracks are required to suppress backgrounds, then we must have a moderately high $\mathcal{P}^{rec}$ so as to place limits on most models.  Finally, for limits in the mass-lifetime plane see the combined plots in \secref{sec:comb}.

%

%% file: Files/Mono-Photons.tex
When the mass splitting between the charged states and the dark matter becomes large enough, they will decay before leaving observable signals in the detector.  The required lifetime is $c\tau \lesssim 1$\;cm, corresponding to $\Delta m \gtrsim 250$\;MeV from \figref{fig:decay}.  This suggests that there is a range, 250\;MeV\;$\lesssim \Delta m \lesssim$~a few GeV, where the decay products will be soft and hard to resolve against the $\gamma\gamma \to$\;hadrons background; $\psi^{q+}\psi^{q-}$ production will be indistinguishable from (invisible) dark matter production.  This extends to the maximal possible mass splitting that can be generated by the non-renormalisable operators of \modeqsref{eq:d7split} and~\eqref{eq:d6split}.  The generic search strategy for invisible final states is to look for production associated with a single energetic object.  For lepton colliders such as CLIC, the usual example of this strategy is a monophoton, a single hard photon together with no other energetic objects in the detector, for the same reasons as in the previous section.

Our dominant signal production process is the same as in the previous section: $e^+e^- \to Z^{\ast}/\gamma^{\ast}\to\psi^{q+}\psi^{q-}$, plus a photon from initial or final state radiation.  This suggests a similar scaling in the signal cross sections, namely triplet\,:\,quintet\,:\,septet close to (but a bit below) $1:17:98$.  We also have non-negligible production through from $e^\pm\gamma \to W^{\ast} \to \chi^{q\pm} \chi^{(q-1)\mp}$ and, for the Higgsino-like doublet, direct production of the dark matter itself, $e^+e^- \to Z^\ast \to\chi\chi$; all of these are included.

\begin{table}
	\centering
	\begin{tabular}{|c|c||c|c|}
		\hline
		Background & Cross Section (ab) & Background & Cross Section (ab) \\
		\hline
		$\nu\nu\gamma$ & $1.9 \times 10^6$ & $e^\pm\gamma$ & $1.5 \times 10^6$ \\
		$\nu\nu\gamma\gamma$ & $5.9 \times 10^4$ & $e^\pm\gamma\gamma$ & $1.8 \times 10^5$ \\
		$\nu\nu\gamma\gamma\gamma$ & 742 & $e^\pm\gamma\gamma\gamma$ & $3.1 \times 10^3$ \\
		$e^+e^-\gamma$ & $1.1 \times 10^6$ & $\mu^+\mu^-\gamma$ & $9.9 \times 10^4$ \\
		$e^+e^-\gamma\gamma$ & $6.4 \times 10^4$ & $\mu^+\mu^-\gamma\gamma$ & $2.6 \times 10^3$ \\
		$e^+e^-\gamma\gamma\gamma$ & 409 & $\mu^+\mu^-\gamma\gamma\gamma$ & 23 \\
		\hline
	\end{tabular}
	\caption{Major backgrounds to mono-photon searches that we consider.  The cross sections (averaged over polarisations) are given for relatively mild cuts: $p_T^\gamma > 25$\;GeV, $\lvert\cos\theta^\gamma\rvert < 0.99$, and a veto on any other particles in the detector volume with $p_T > 10$\;GeV.}\label{tab:bkg}
\end{table}

There are a large number of potentially relevant backgrounds.  Most importantly, there is an irreducible background $e^+e^- \to \nu\bar{\nu}\gamma$ with a pb-scale cross section.  We also consider a number of reducible backgrounds, including $e^+e^- \to e^+e^-\gamma$ and $e^\pm\gamma \to e^\pm\gamma$, listed in \tabref{tab:bkg}.  These are all relatively simple electrodynamic processes, so we assume a small systematic uncertainty, which we vary in the range $\epsilon_{sys} = \{0, 0.5\%, 1\%\}$.  An important complication for our model is that the irreducible background is generated through the same $SU(2)_L$ coupling as the signal.  Indeed, there are two major contributions: the radiative return process $e^+e^- \to Z\gamma$ followed by invisible $Z$ decay, and $t$-channel $W$ exchange.  In most monophoton studies at linear colliders, the latter is suppressed by choosing the initial beam polarisation $(P(e^-), P(e^+)) = (+,-)$, which suppresses the electron-$W$ coupling.  Here, doing so will equally suppress the signal cross section, which will limit the reach of this search.  Indeed, given that the \emph{reducible} backgrounds are approximately helicity-independent, it is sometimes optimal to consider $(P(e^-), P(e^+)) = (-,+)$ maximising the signal cross section.

The combination of a large background and a very simple final state (defined by a single four-vector) forces a slightly different approach to placing cuts.  Our choice of the \emph{type} of cuts we impose is based on the properties of the background, but we vary the actual value of the cut with the DM mass so as to maximise the signal significance.  We suppress reducible backgrounds by vetoing all events with more than one reconstructed photon, or any leptons with $p_T > 25$~GeV.  We also assume that when a photon overlaps with a charged track, it will not be reconstructed as a photon; and when two photons overlap, they will not pass experimental purity cuts.  This (small) background contribution is also rejected.  We apply three cuts on the final state photon kinematics:
\begin{itemize}
	\item An upper bound on the energy, $E_\gamma < E^{cut}$;
	\item A lower bound on the transverse momentum, $p_T^\gamma > p_T^{cut}$;
	\item And an angular cut, $\lvert \cos\theta_\gamma\rvert < \cos\theta^{cut}$, implemented as a cut on the rapidity $\lvert\eta_\gamma\rvert < \eta^{cut}$.
\end{itemize}
The first cut is aimed at the radiative return process.  As a two-body final state, the photon energy in the collision frame is predicted to be
\begin{equation}
	E_\gamma^{RR} = \frac{\sqrt{\hat{s}}}{2} \, \biggl( 1 - \frac{m_Z^2}{\hat{s}} \biggr) \,,
\end{equation}
where $\hat{s}$ is the centre of mass energy.  When the collision is at rest in the lab frame and $\sqrt{\hat{s}} = 3$\;TeV, this corresponds to a photon energy $E_\gamma = 1.498$\;TeV.  The large beamsstrahlung effects expected at CLIC mean that a significant fraction of collisions occur at lower energies and in a boosted frame; however, we still expect this background to be peaked at high photon energies.  By way of comparison, the energy of the photon in the signal process is 
\begin{equation}
	E_\gamma^{sig} \leq \frac{\sqrt{\hat{s}}}{2} \, \biggl( 1 - \frac{4m_{DM}^2}{\hat{s}} \biggr) \,.
\label{eq:Esigupper}\end{equation}
As the DM mass increases more of the signal is concentrated at lower photon energies, and can use a more stringent energy cut without rejecting any signal.

The other two cuts are aimed at the $W$-exchange contribution to the irreducible background, as well as the reducible backgrounds.  Background photons from the former are softer and more forward due to it being $t$-channel while the signal is $s$-channel, so both cuts are effective in improving sensitivity.  The $p_T$ cut is also very important in reducing the irreducible backgrounds, which only contribute if the additional photons and leptons are either soft or collinear.  Demanding they recoil against a hard photon can make this impossible.

\begin{figure}
	\centering
	\includegraphics[width=0.32\textwidth]{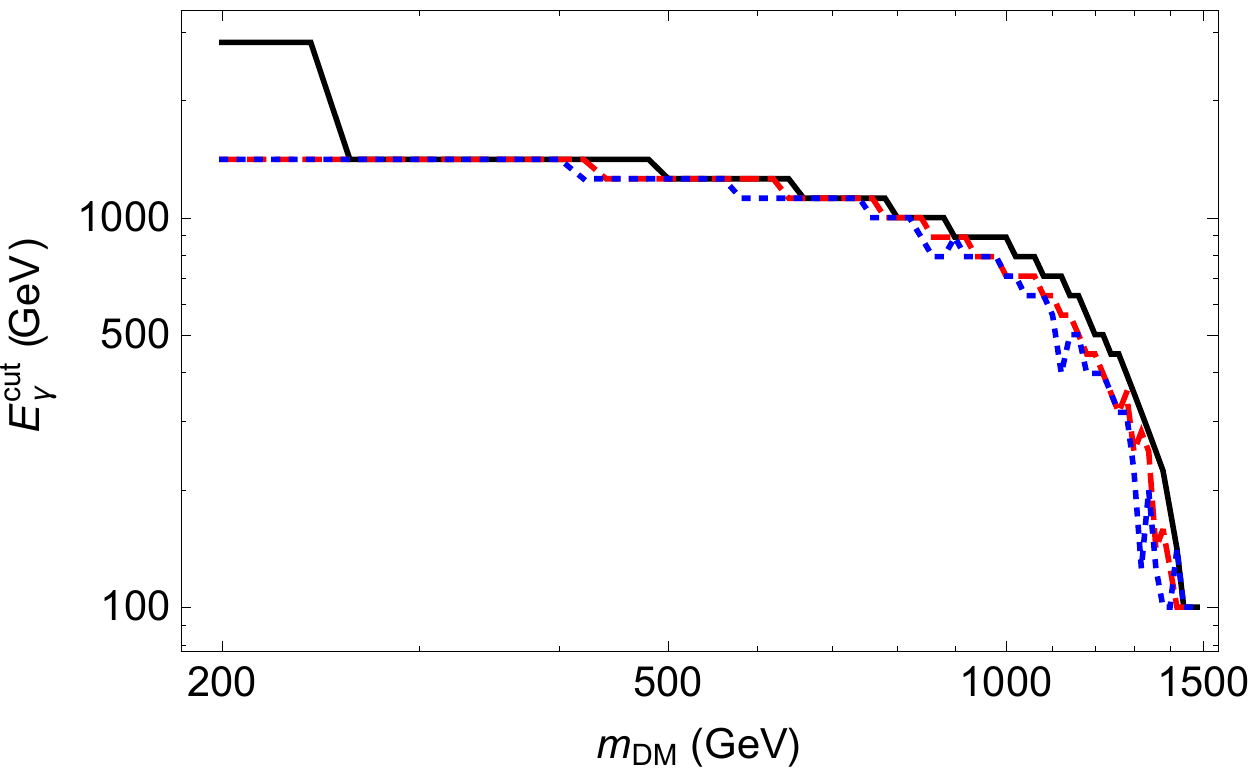}\
	\includegraphics[width=0.32\textwidth]{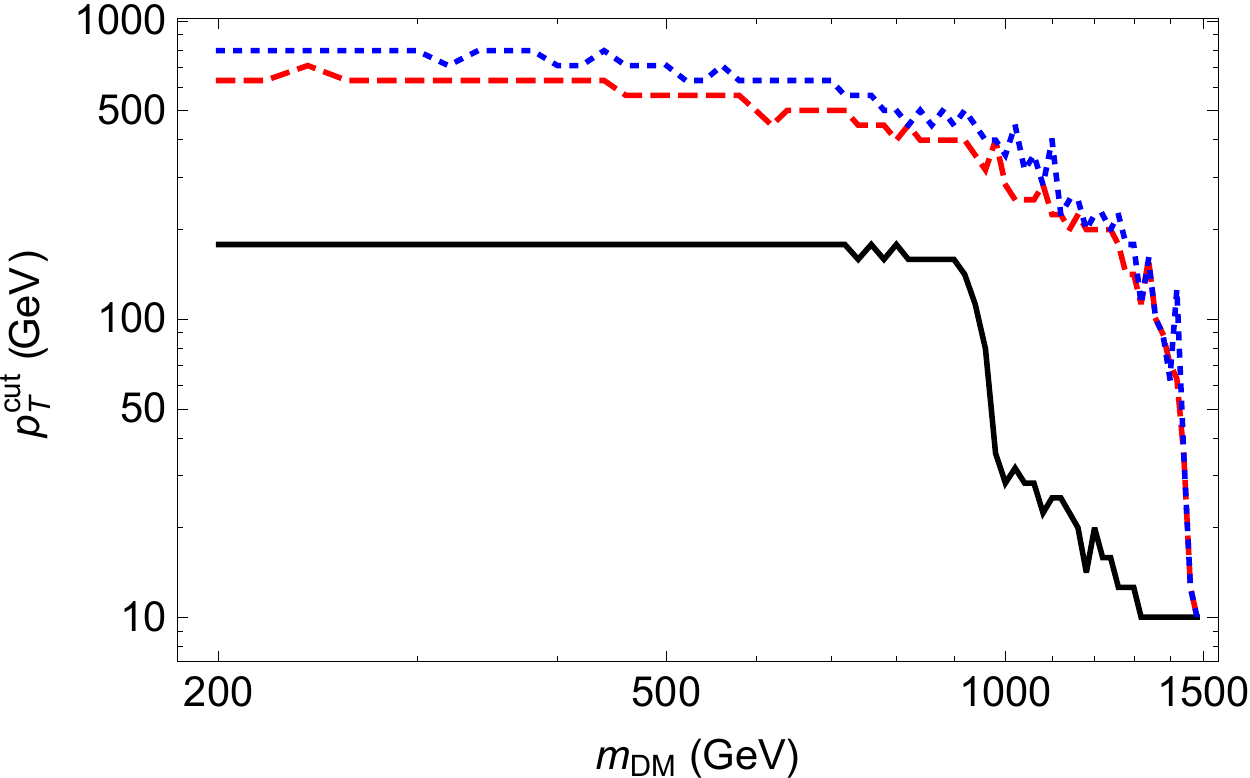}\
	\includegraphics[width=0.32\textwidth]{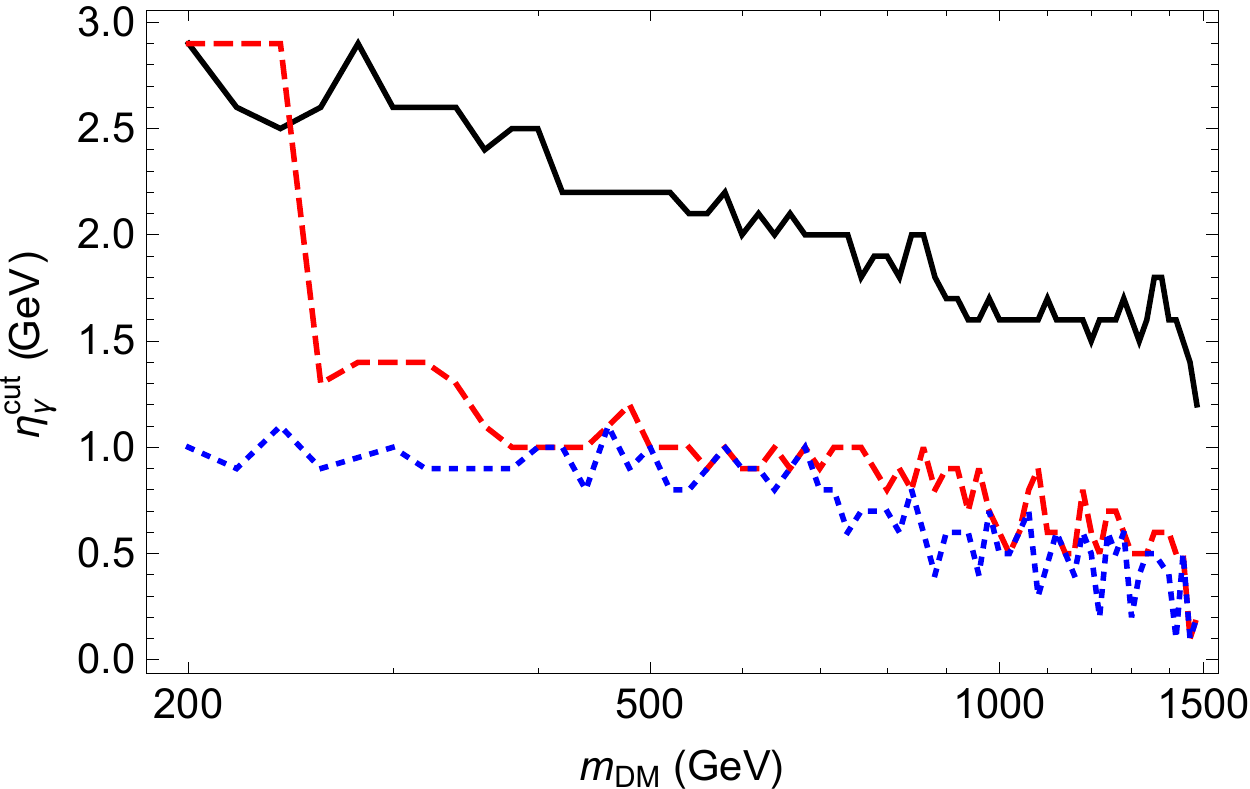}\\
	\includegraphics[width=0.32\textwidth]{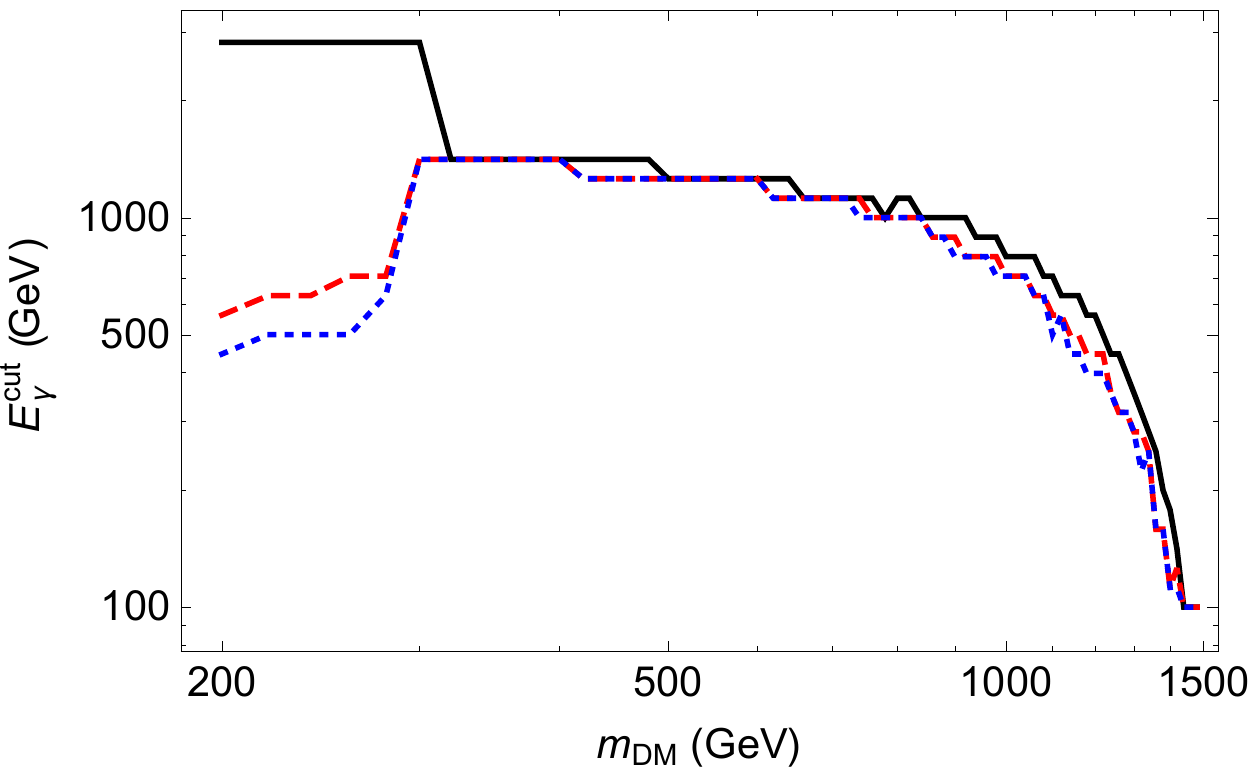}\
	\includegraphics[width=0.32\textwidth]{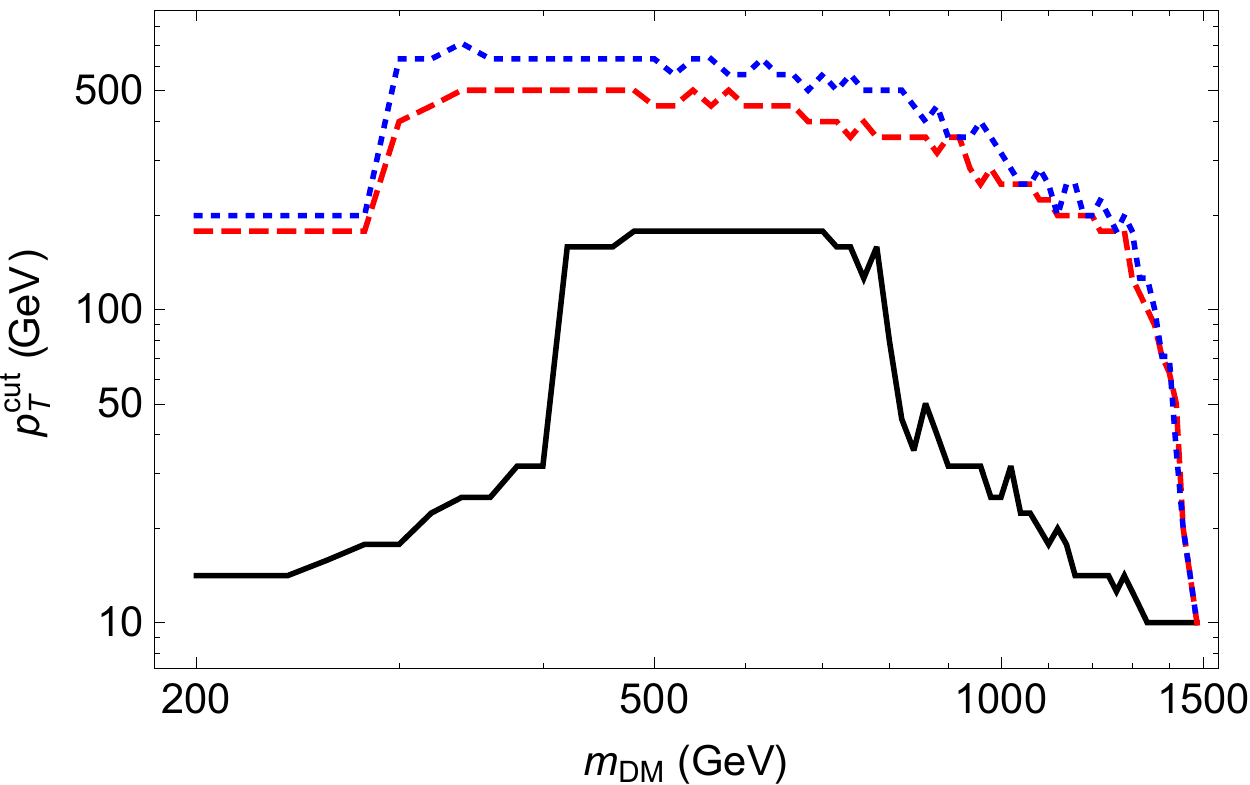}\
	\includegraphics[width=0.32\textwidth]{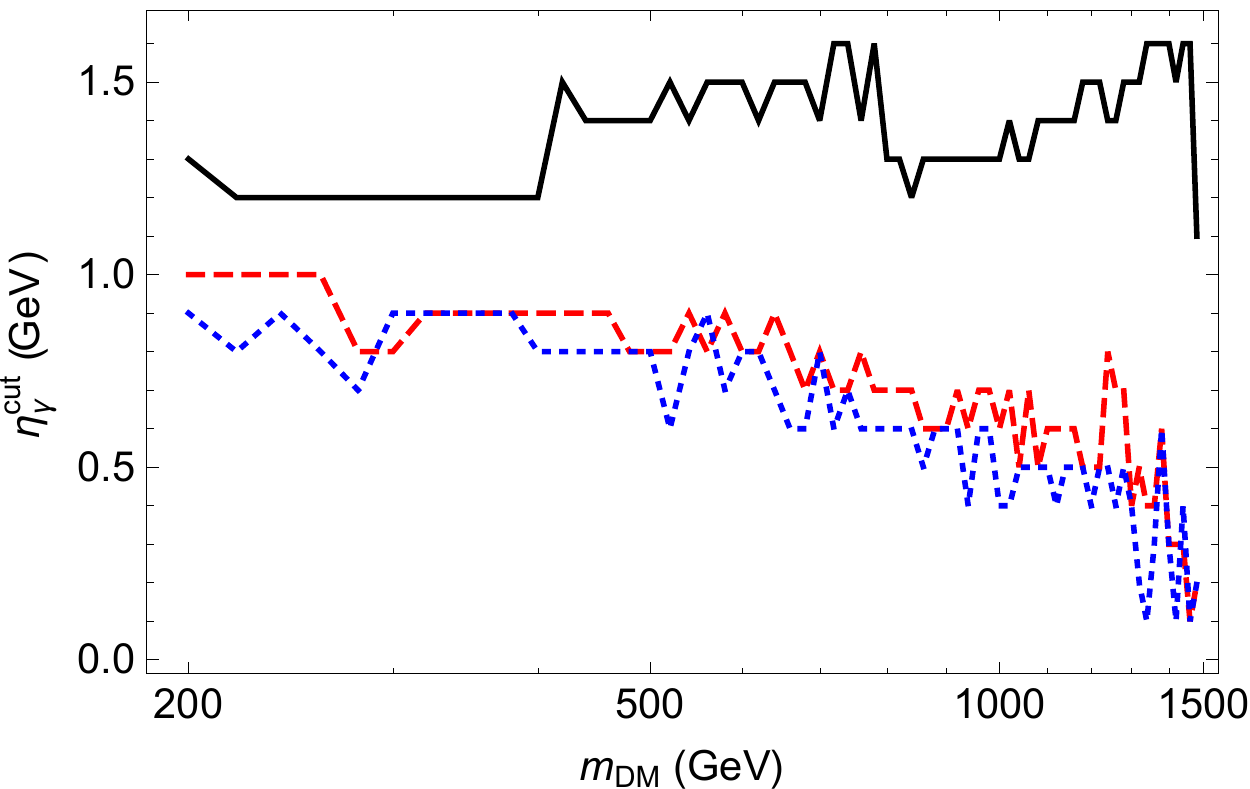}
	\caption{Cuts that optimise significance for a fermion triplet (top) and quintet (bottom).  The black solid (red dashed, blue dotted) lines are for 0\% (0.5\%, 1\%) systematic uncertainties, when we include the photon smearing and resolution effects, and for the beam polarisation $(P(e^-), P(e^+)) = (-,+)$.  The optimal cuts for other multiplets and polarisations are similar.}\label{fig:cuts}
\end{figure}

We allow our three cuts to vary in the ranges
\begin{equation}
	p_T^{cut} \in [25, 1000) \text{ GeV}, \qquad E^{cut} \in [ p_T^{cut}, 1.5 \text{ TeV})\,, \qquad \eta^{cut} \in [0.1, 3) \,.
\end{equation}
We optimise the cuts to maximise the signal significance for each mass point considered; and also separately for the different beam polarisations, detector approximations, and systematic uncertainties.  We assume the full design integrated luminosity of 2\,ab$^{-1}$, weighted as in \tabref{tab:lumi}.  The resultant cuts are illustrated in \figref{fig:cuts} for the fermion triplet and quintet; the results for the scalars and other multiplets are qualitatively similar.  The behaviour of these cuts as a function of mass can be understood as follows:
\begin{itemize}
	\item The energy cut $E^{cut}$ is over 1\;TeV for the lightest DM we consider, and monotonically decreases with increasing $m_\chi$.  This is due to the effect mentioned above that as $m_\chi$ increases, the upper limit on the signal photon energy \modeqref{eq:Esigupper} decreases and the stronger cut only removes background.
	\item The momentum cut exhibits a similar dependence on mass.  At low masses when the energy cut is mild, a large $p_T^{cut}$ is needed to suppress the $t$-channel and reducible backgrounds.  As $m_\chi$ increases, the energy upper limit becomes more important in cutting the backgrounds.   $p_T^{cut}$ can and \emph{must} decrease, since unless $p_T^{cut} < E^{cut}$ no events will pass our cuts.
	\item The rapidity cut's role in suppressing the $t$-channel backgrounds results in $\eta^{cut}$ decreasing at high masses to compensate for the weaker $p_T$ cut.
\end{itemize}

We also see a clear dependence in the cuts on the systematic uncertainties.  In the ideal case $\epsilon_{sys} = 0\%$, looser cuts enhance the signal \emph{and} reduce the statistical uncertainty on the background.  This leads to the feature seen in \figref{fig:cuts} where, at fermion quintet low masses, the optimal $p_T$ is very low.  However, when we consider more realistic values of the systematic uncertainty it dominates the statistical effect, forcing us to make very severe cuts in an attempt to make the background as small as possible.  This also leads to the feature observable in \figref{fig:cuts} where the cuts in the ideal case exhibit some dependence on the DM representation, but for realistic $\epsilon_{sys}$ are very similar because of the need to suppress the background.

\begin{figure}
	\centering
	\includegraphics[width=0.45\textwidth]{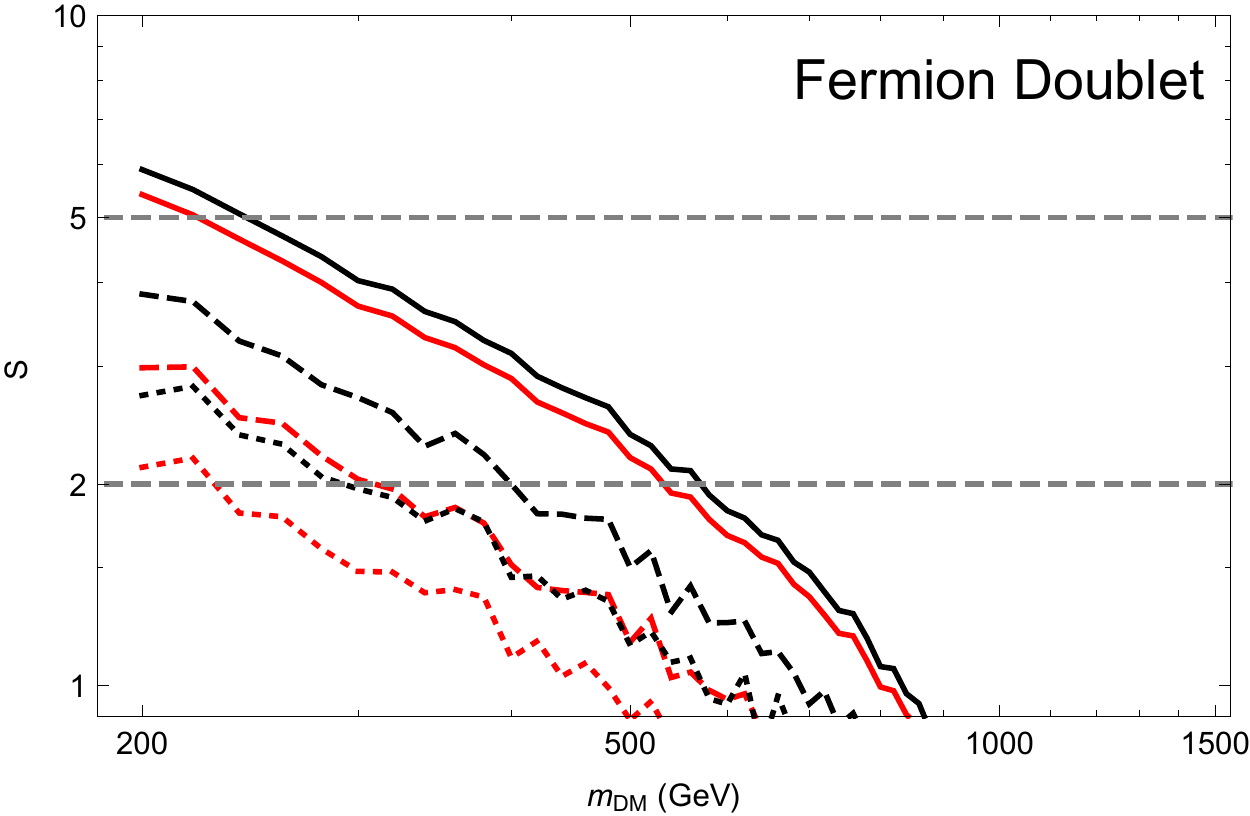}\quad
	\includegraphics[width=0.45\textwidth]{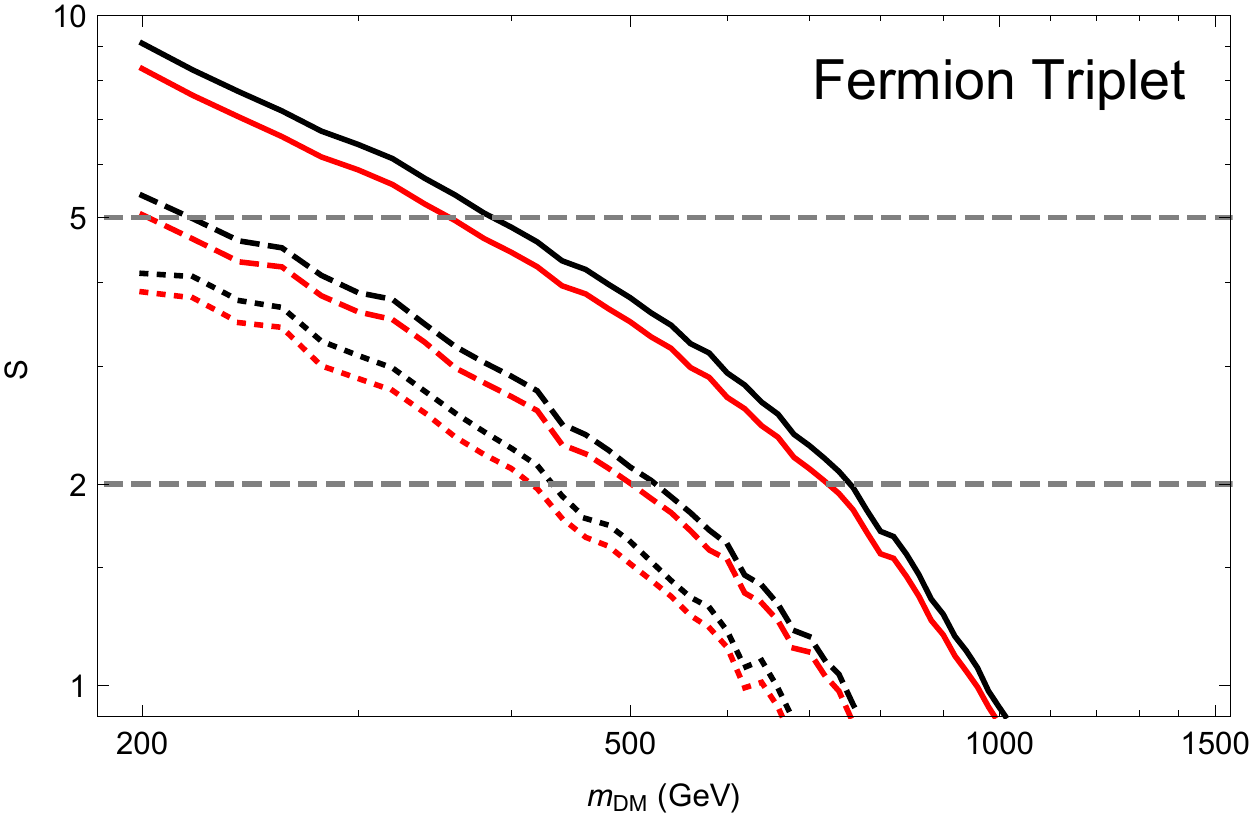}\\
	\includegraphics[width=0.45\textwidth]{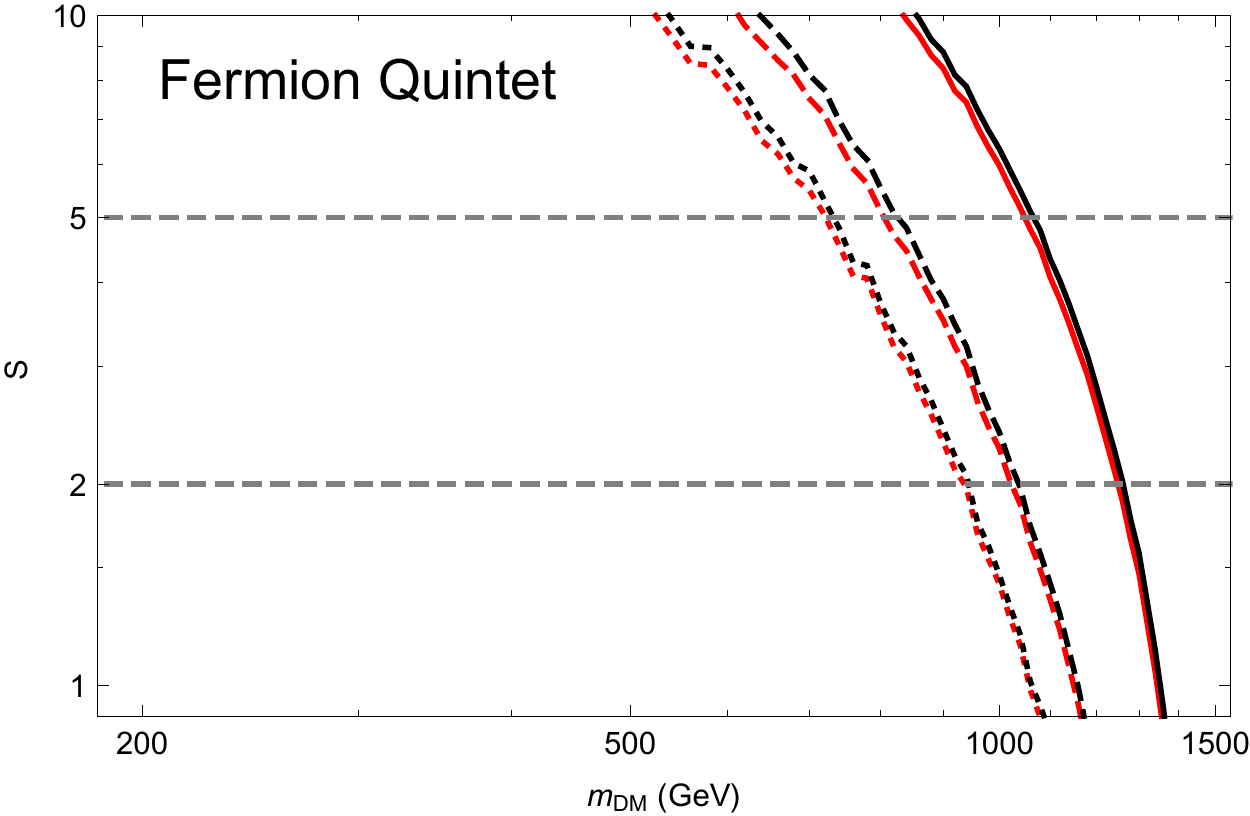}\quad
	\includegraphics[width=0.45\textwidth]{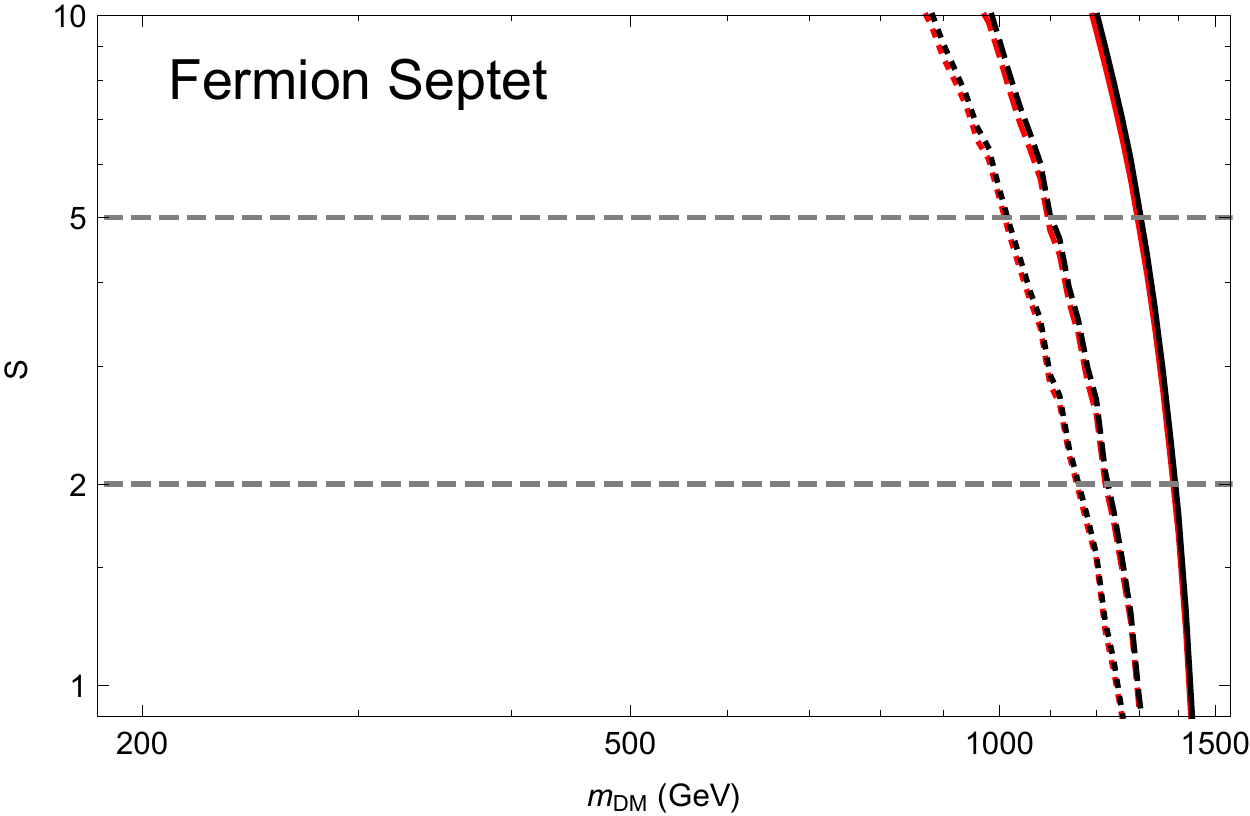}
	\caption{Significances for fermionic EMDM for 2\;ab$^{-1}$ of CLIC data in the mono-photon channel.  Top row: Complex doublet (left) and real triplet (right).  Bottom row: real quintet (left) and real septet (right).  The black lines are for limits using parton-level observables; the red for those with the detector simulation as described in the text.  Solid (dashed, dotted) lines are for 0\% (0.5\%, 1\%) systematic uncertainties.  The grey dashed horizontal lines show the 5$\sigma$-discovery and 95\%-exclusion points.}\label{fig:monoPlimF}
\end{figure}

\begin{figure}
	\centering
	\includegraphics[width=0.31\textwidth]{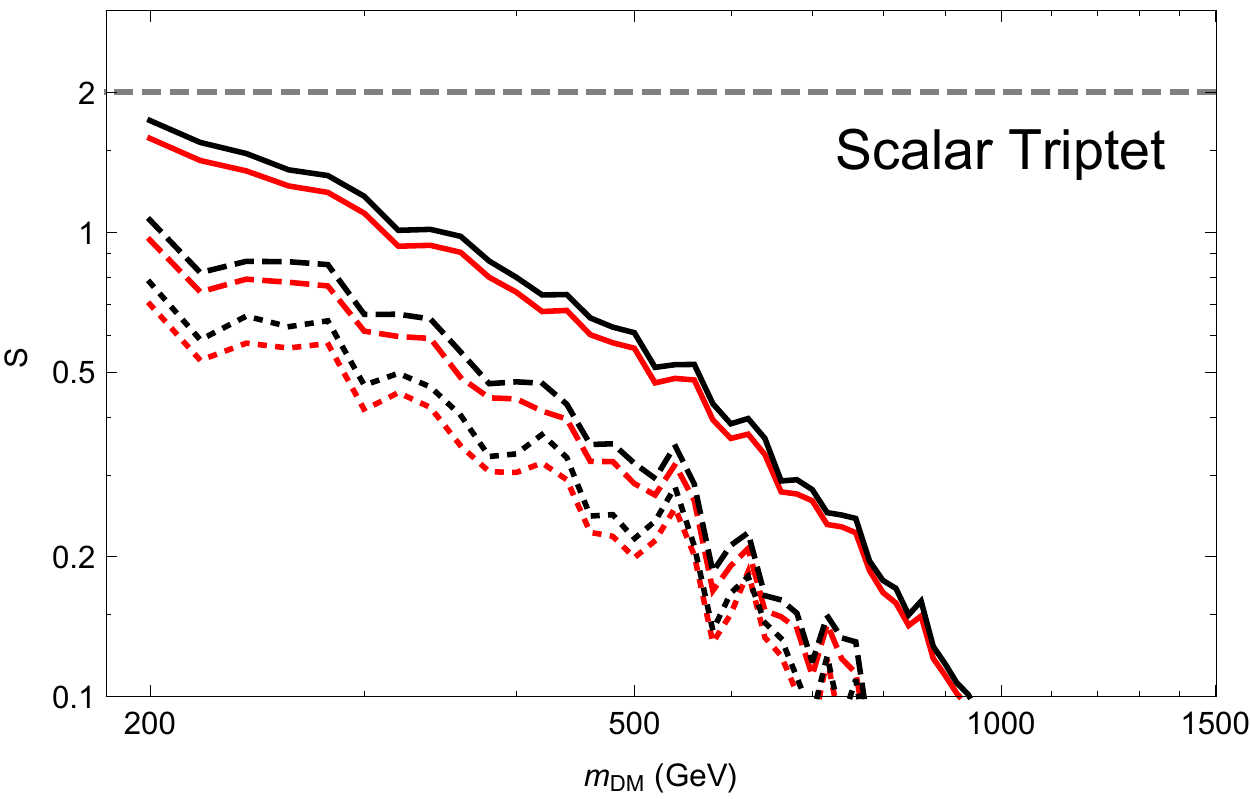}\
	\includegraphics[width=0.31\textwidth]{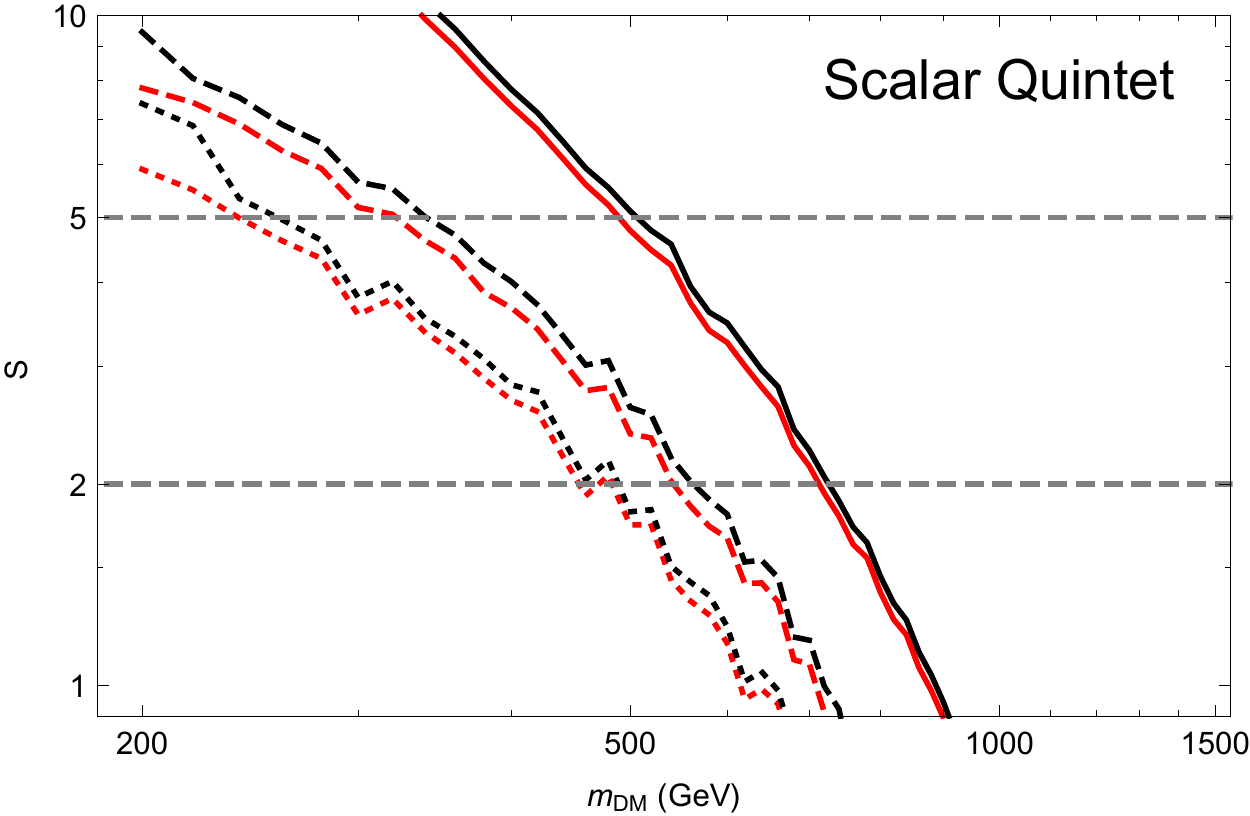}\
	\includegraphics[width=0.31\textwidth]{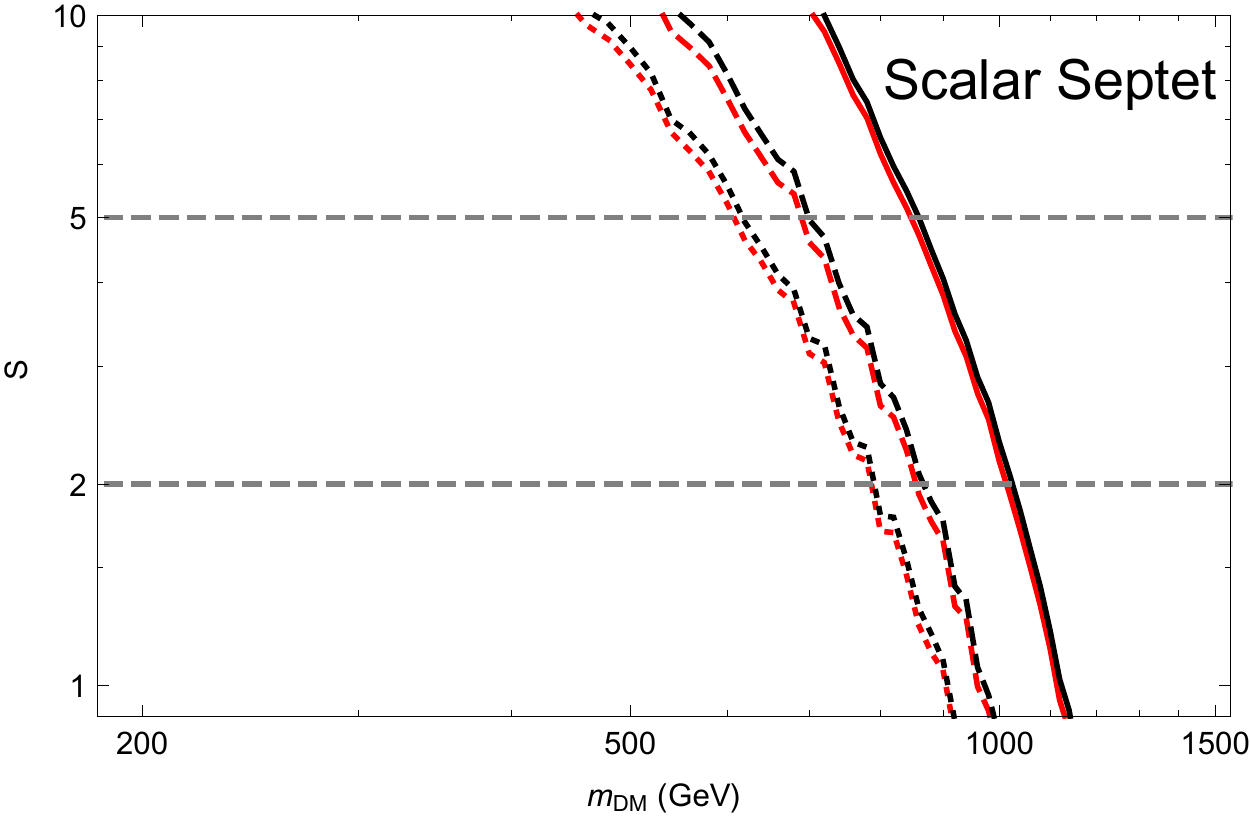}
	\caption{Significances for scalar EMDM for 2\;ab$^{-1}$ of CLIC data in the mono-photon channel for, from left to right, the real scalar triplet, quintet, and septet respectively.  The notation is as in \figref{fig:monoPlimF}.}\label{fig:monoPlimS}
\end{figure}

Using the optimal cuts derived as described, we can compute the expected discovery and exclusion reaches.  We checked whether superior limits derive by including all data, or only a subset of beam polarisations.  Due to the signal and irreducible backgrounds having the same dependence on polarisation, and the severe cuts suppressing the reducible backgrounds, the significance is typically maximised by including all data.  We show the expected significance as a function of DM mass for our different multiplets in \figrefs{fig:monoPlimF} and~\ref{fig:monoPlimS}.  We also mark the 5$\sigma$-discovery and 95\%-exclusion points.  The limits are stronger for larger and for fermionic multiplets, due to the enhanced cross sections.  Non-zero systematic uncertainties suppress the expected reach by a few hundred GeV.  This effect is relatively less important for the larger multiplets because the larger signals allow stronger cuts, making the statistical uncertainty relatively more important.  In the most studied cases of the Higgsino-like doublet and Wino-like triplet models, prospective exclusions in this channel are relatively weak, approximately 300\;GeV and 500\;GeV respectively.  The larger fermion multiplets can be discovered closer to the kinematic limit $m_{DM} \sim 1$--1.5\;TeV.  Lastly, there is no expected sensitivity to the scalar triplet in this channel, but bounds on the scalar quintet (septet) are expected to be comparable to those for the fermion triplet (quintet).

%

%% file: Files/comb.tex
\begin{figure}
	\centering
	\includegraphics[width=0.45\textwidth]{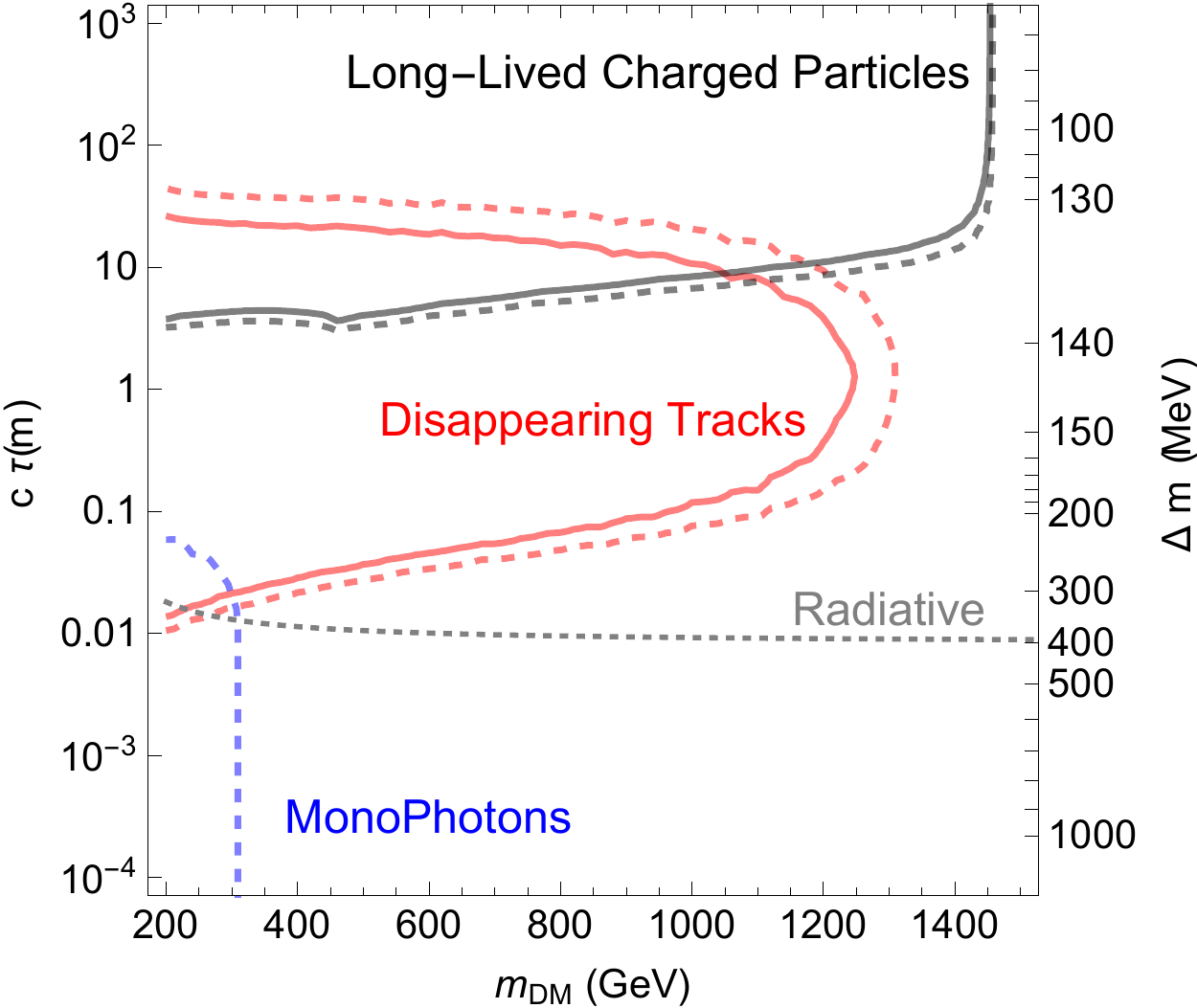}\
	\includegraphics[width=0.45\textwidth]{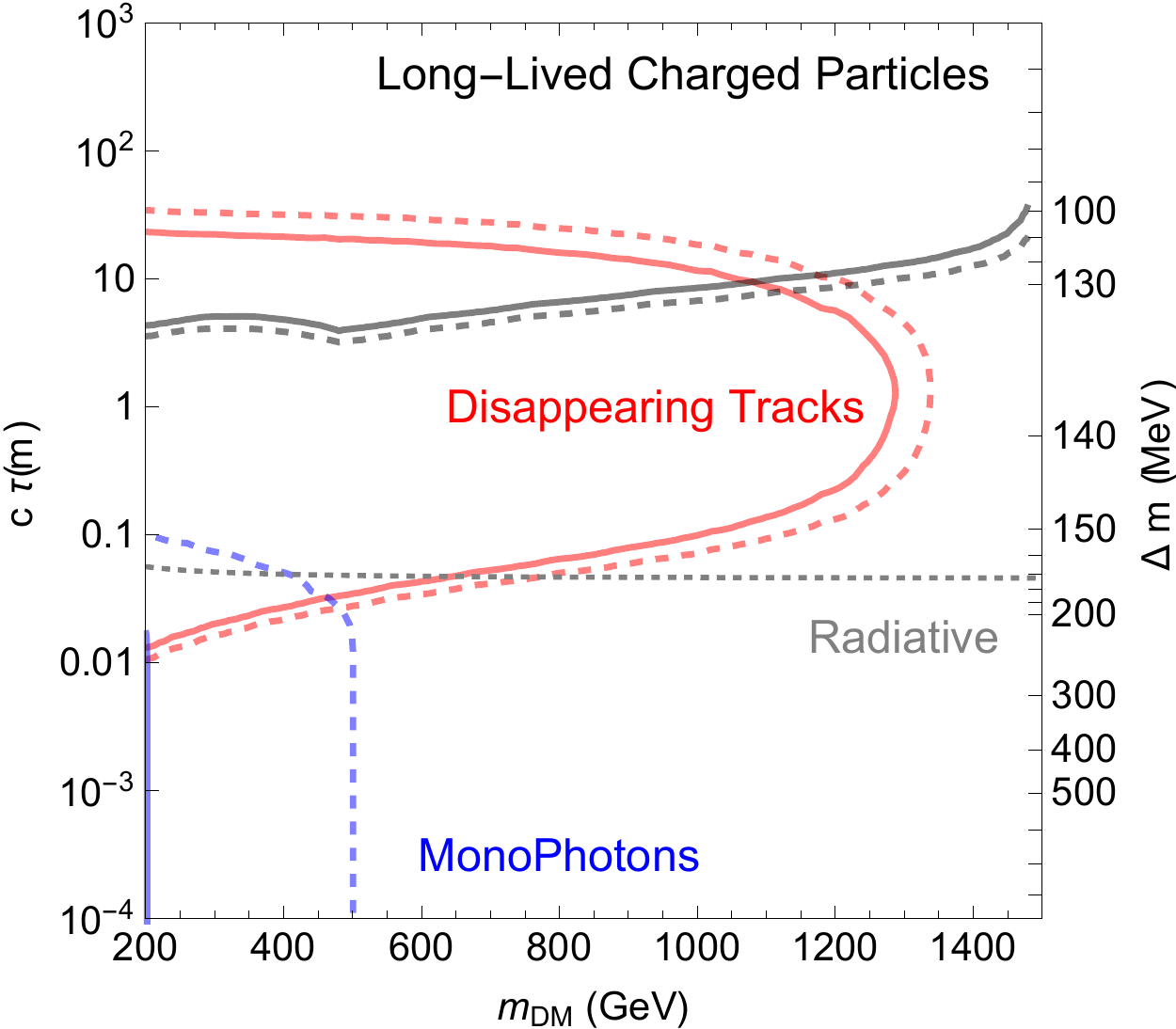}\\
	\includegraphics[width=0.45\textwidth]{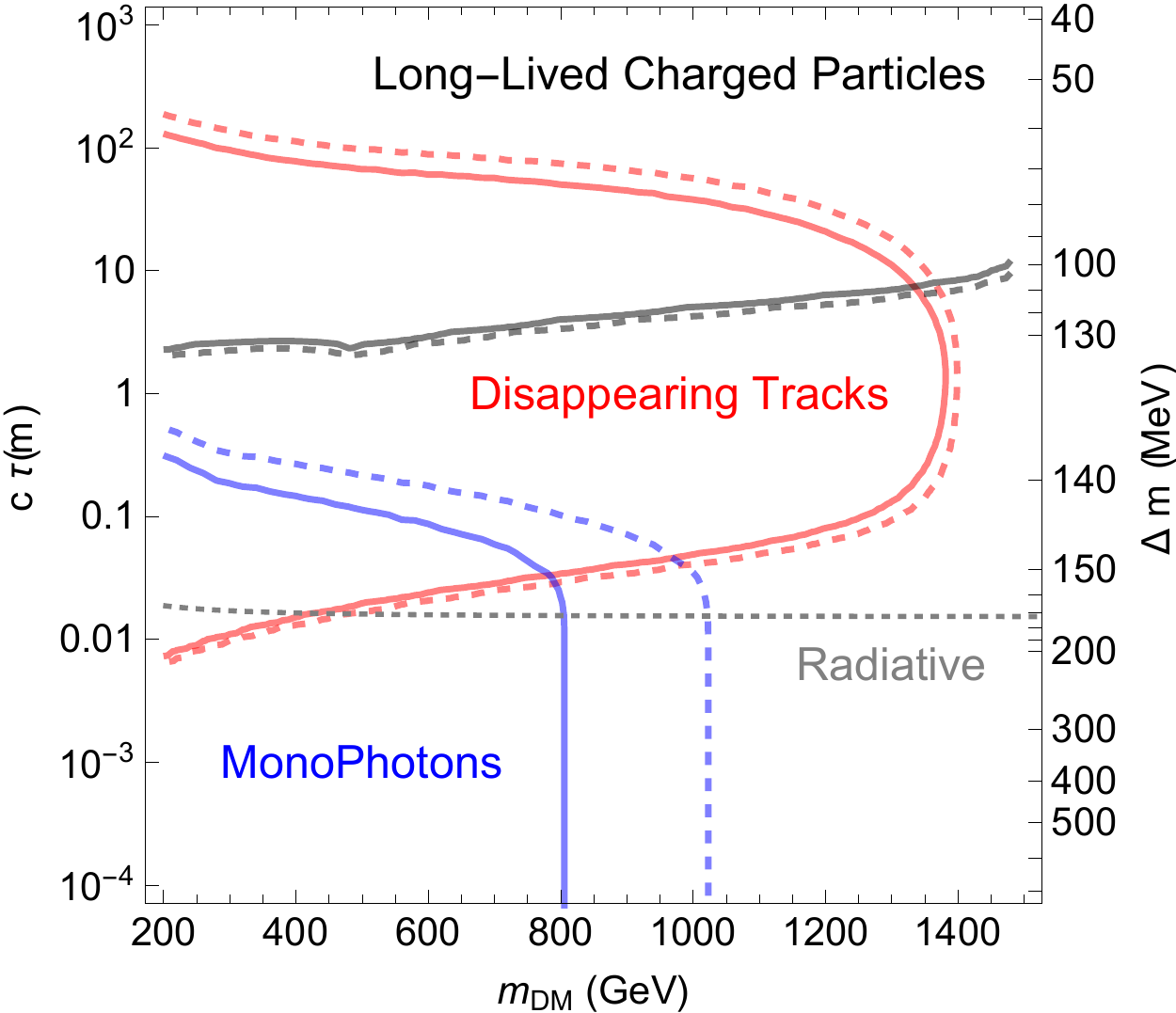}\
	\includegraphics[width=0.45\textwidth]{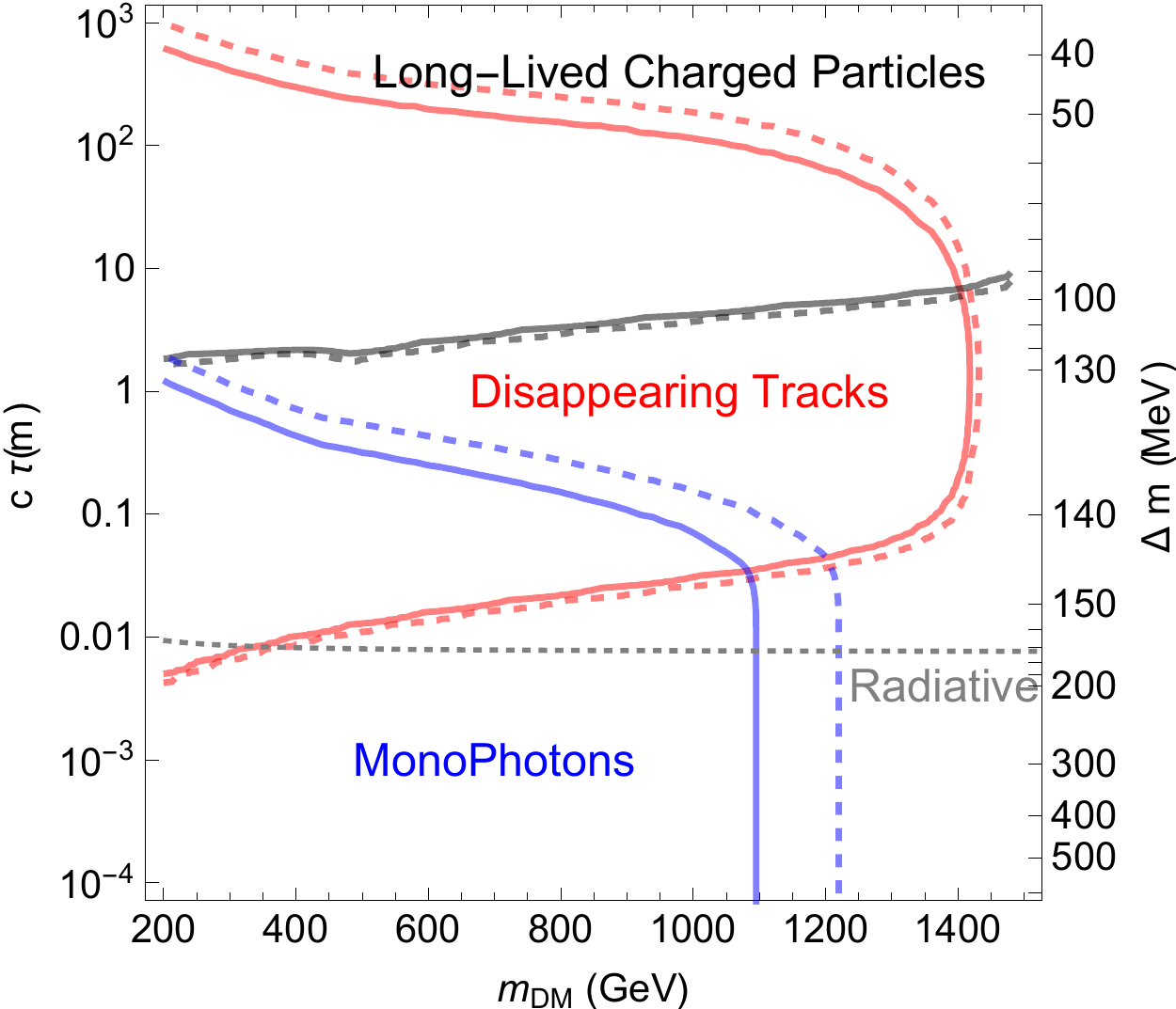}
	\caption{Combined exclusion and discovery plots for fermionic EMDM in the mass-lifetime plane.  Top row: Complex doublet (left) and real triplet (right).  Bottom row: real quintet (left) and real septet (right).  The different contours describe the searches from \secrefs{sec:llcp}, \ref{sec:dtrack} and \ref{sec:monophotons} as labelled.  The line marked `Radiative' shows the mass splitting generated purely from radiative corrections.  See the text for additional details.}\label{fig:combFerm}
\end{figure}
\begin{figure}
	\centering
	\includegraphics[width=0.31\textwidth]{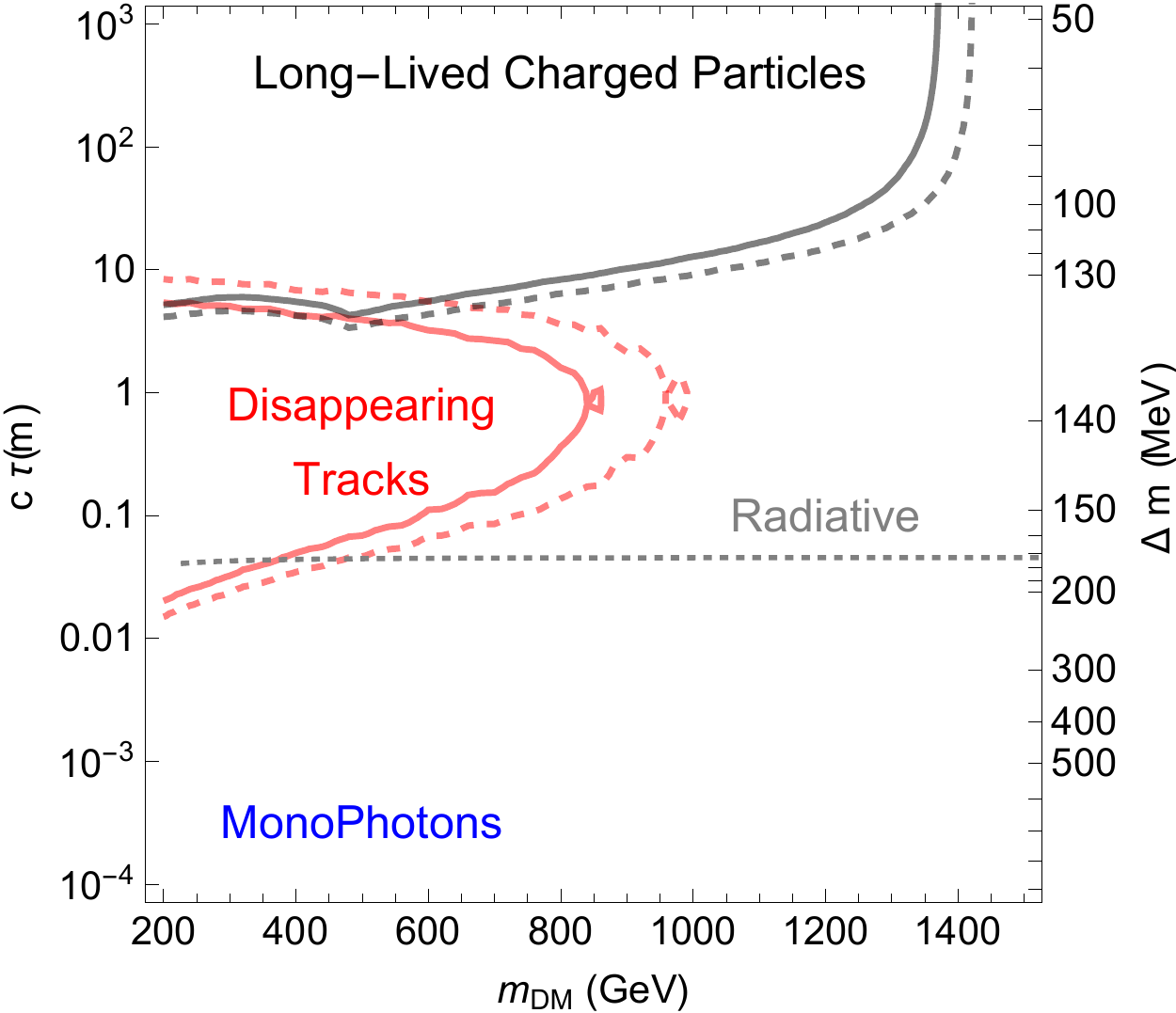}\
	\includegraphics[width=0.31\textwidth]{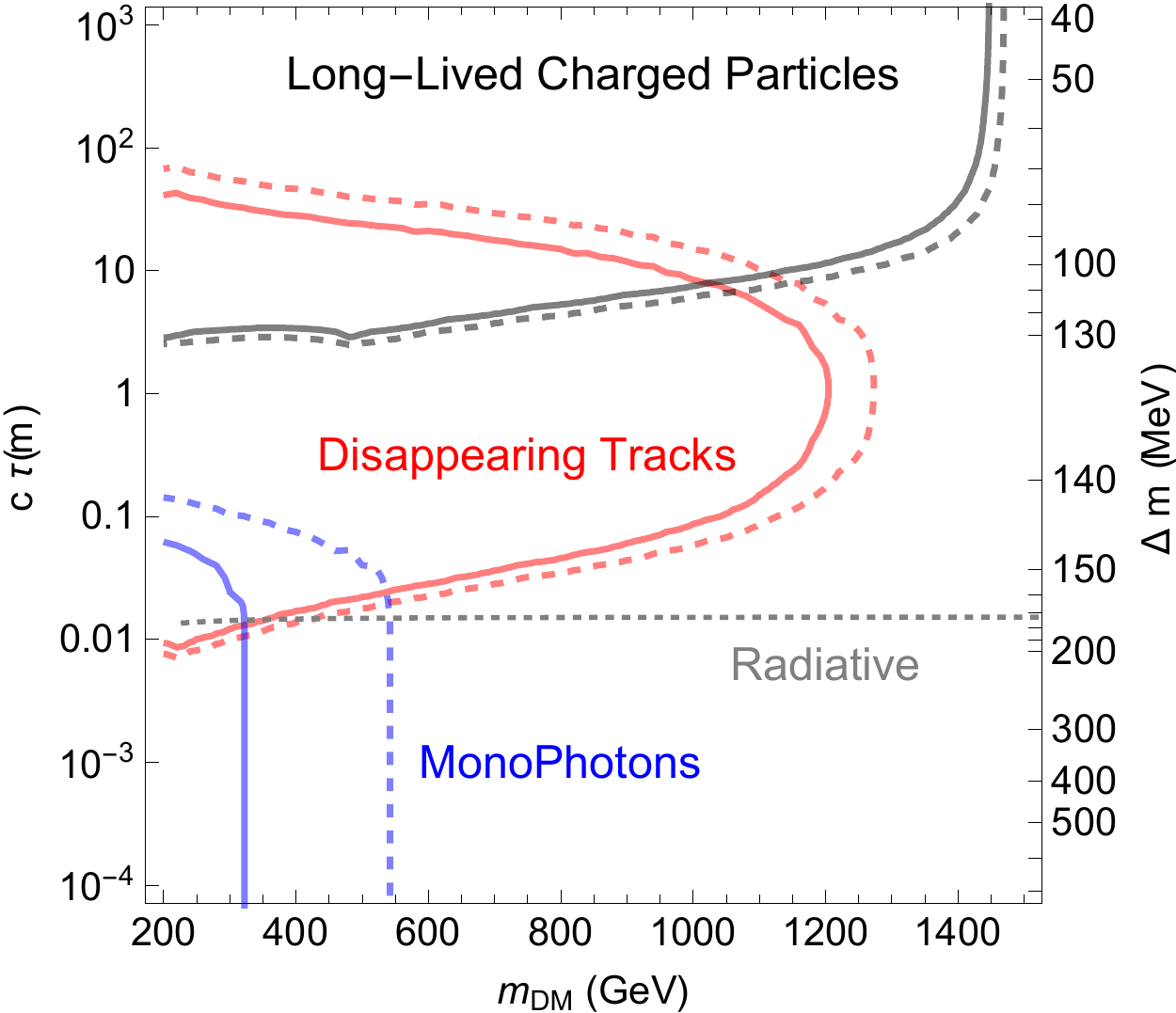}\
	\includegraphics[width=0.31\textwidth]{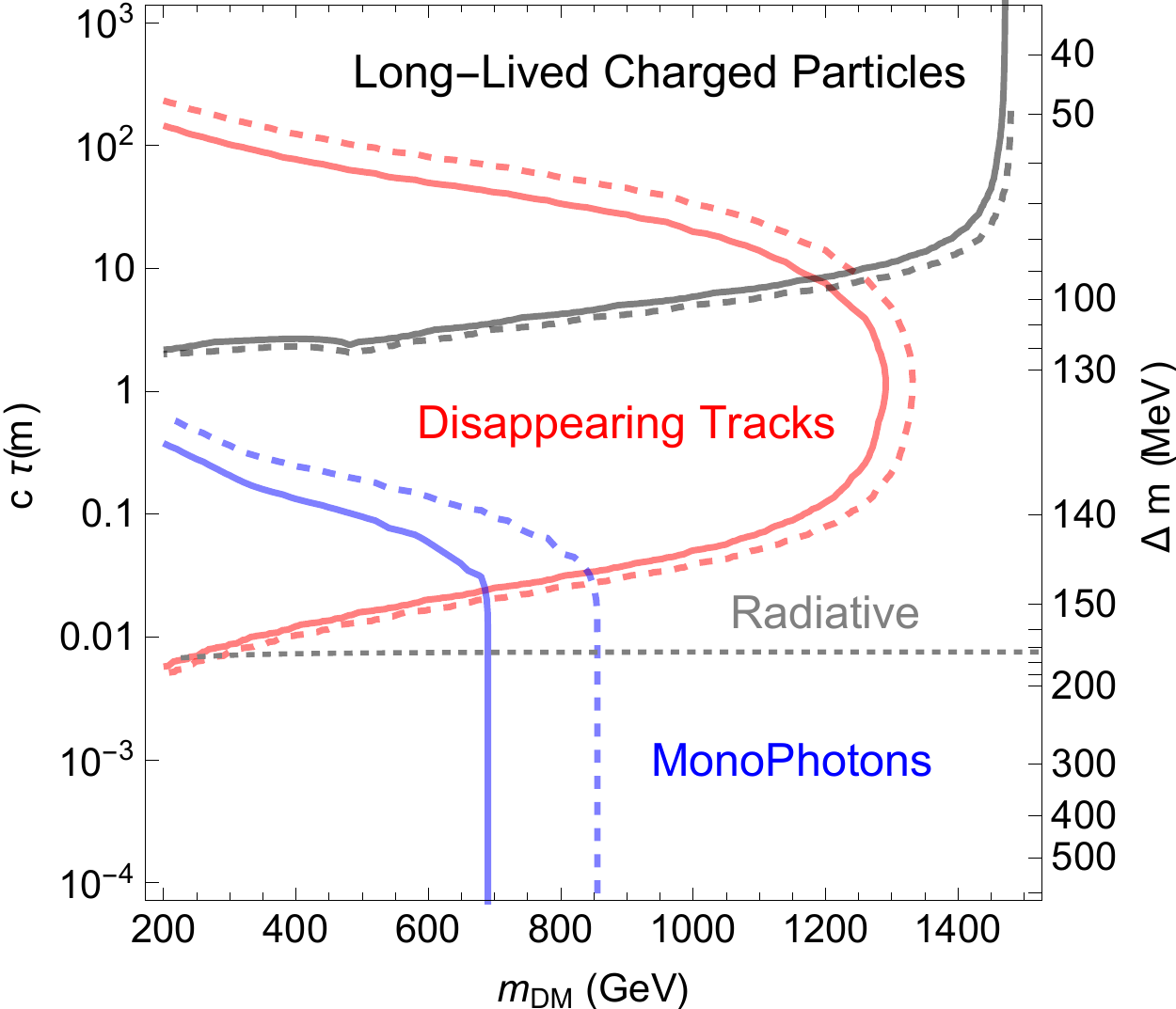}
	\caption{Combined exclusion and discovery plots for scalar EMDM in the mass-lifetime plane for, from left to right, the real scalar triplet, quintet, and septet respectively.  The notation is as in \figref{fig:combFerm}.}\label{fig:combSca}
\end{figure}

In this section, we combine the results from the previous sections to find the full discovery and exclusion reach at CLIC from direct searches.  We show the results for our four fermion models in \figref{fig:combFerm}, and the three scalar models in \figref{fig:combSca}.  The solid (dashed) lines in these plots show 5$\sigma$ discovery (95\% exclusion) contours.  We make the following specific choices from those discussed in the previous sections:
\begin{itemize}
	\item For LLCP searches (shown in black), we include detector reconstruction effects, a 50\% systematic uncertainty, and use the two-LLCP strategy as discussed in \secref{sec:llcp}.
	\item For disappearing track searches (red), we demand two such tracks with a reconstruction efficiency of $\mathcal{P}^{rec} = 30\%$ as outlined in \secref{sec:dtrack}.  Additionally, since we do not have a significance estimate we demand 10 (5) events for discovery (exclusion).
	\item For monophoton searches (blue), we assume a 0.5\% systematic uncertainty and include our detector reconstruction effects, see \secref{sec:monophotons}.
\end{itemize}
We plot our results in the $m_\chi$-$c\tau$ plane, where $c\tau$ is the lifetime of the singly-charged state.  Since the lifetime is determined by the mass splitting $\Delta m_1$, we show that on the right-hand vertical axis.  We also show the line that corresponds to a purely-radiative mass splitting in grey.

The limits we derive are stronger for fermions and for larger multiplets, as these states have bigger production cross sections.  For $c\tau \gtrsim 10$\;m, LLCP searches exclude states (nearly) all the way to the kinematic limit, $m_\chi > 1.5$\;TeV.  At smaller lifetimes, the charged states decay before leaving the collider and disappearing track searches become relevant.  Our demand for a hard photon limits the maximum reach to $\sim 1.4$\;TeV, which is achieved for larger multiplets.  As discussed in \secref{sec:dtrack}, including the photon is a conservative choice and it may be possible to improve on this.  We can see in \figrefs{fig:combFerm} and~\ref{fig:combSca} that these limits are maximal for lifetimes around 1\;m.  Longer-lived states survive into the muon chambers, and more closely resemble LLCPs; while as the lifetime decreases, the charged states decay before travelling far enough into the tracking system.  The latter behaviour is sensitive to the details of the ultimate detector design including the spacing and the number of hits required to identify a track, and so the exact position of the lower edge to the excluded region is likely to shift.  Finally, as the lifetime drops below a few cm, the charged states decay within the beam pipe leaving no easily observable decay products.  With no direct signal of the dark matter multiplet, we set limits using monophoton searches instead.  These are much weaker than the other searches we consider due to the large backgrounds.  We discuss possible ways to improve prospects for these lifetimes in the following section.

There are two results of particular interest.  The first are the prospects when the mass splitting is purely radiative, \emph{i.e.} the operators of \modeqsref{eq:d5split}, \eqref{eq:d7split} and \eqref{eq:d6split} are negligible.  These can be easily extracted from \figrefs{fig:combFerm} and~\ref{fig:combSca} and we list them in \tabref{tab:radexc} for convenience.  The most relevant signals here are disappearing tracks for the two triplets, and monophotons for all other models.  The other natural question is what bounds can be placed when the DM has the correct thermal relic density from gauge interactions (alone).  This picks out a specific mass value for each multiplet.  The majority are too heavy to be directly searched for at CLIC; the Higgsino-like fermion doublet is the sole exception, for which the relevant mass is $m_\chi \approx 1$\;TeV.  The combination of LLCP and disappearing tracks searches exclude mass splittings $\Delta m_1 \lesssim 230$\;MeV, compared to the radiative splitting $\approx 400$\;MeV.  In terms of the dimension-5 operator of \modeqref{eq:d5split}, this corresponds to $\Lambda/\lvert c_{5\psi}\rvert \gtrsim 110$\;TeV or $c_{5\psi} > 0$.

\begin{table}
	\centering
	\begin{tabular}{|c||c|c||c|c|}
		\hline
		Multiplet & Fermion & Fermion & Scalar & Scalar \\
		& Exclusion & Discovery & Exclusion & Discovery\\
		\hline
		Doublet & 310 & 230 & n/a & n/a\\
		Triplet & 775 & 600 & 470 & 375 \\
		Quintet & 1025 & 800 & 590 & 375 \\
		Septet & 1220 & 1100 & 850 & 680 \\
		\hline
	\end{tabular}
	\caption{Prospective exclusions and discovery reach in GeV from direct searches at CLIC when the mass splitting is purely radiative.  For the scalar and fermion triplet, these limits derive from disappearing track searches; for the remaining models, they are set by monophotons.}\label{tab:radexc}
\end{table}

For the doublet and triplet models, the constraints we have derived are straightforward.  The LLCP limits extend unchanged to arbitrary small $\Delta m_1$ and long $c\tau$.  For the triplets, the mono-photon constraints extend unchanged to the maximal $\Delta m_1 \sim 1$\;GeV discussed in \secref{sec:models}.  For the Higgsino-like doublet, the existence of \modeqref{eq:d5split} allows much larger mass splittings $\Delta m_1 \gtrsim 10$\;GeV, such that the decay products eventually become sufficiently hard that they can be easily reconstructed.  In this case, the monophoton bounds weaken, but new searches based on these decay products take over.

\begin{figure}
	\centering
	\begin{tikzpicture}[node distance=0.4cm and 0.4cm]
		\coordinate[label=below:{production}] (v1);
		\coordinate[above right = of v1, label=right:{decay}] (v2);
		\draw[fill] circle [radius=2pt];
		\draw[thick,red] circle [radius=1.6];
		\draw[fill] (v2) circle [radius=1pt];
		\draw[fermionnoarrow] (v1) -- (v2);
		\coordinate[above left = of v2] (va);
		\coordinate[above left = of va] (vb);
		\coordinate[above = of vb] (v3);
		\draw[fermionnoarrow] (v2) -- (v3);
	\end{tikzpicture}\qquad\qquad
	\begin{tikzpicture}[node distance=0.2cm and 0.2cm]
		\coordinate (v0);
		\coordinate[right = of v1] (v1a);
		\coordinate[right = of v1a] (v1b);
		\coordinate[right = of v1b] (v1c);
		\coordinate[right = of v1c] (v1d);
		\coordinate[right = of v1d] (v1e);
		\coordinate[right = of v1e, label = below:{beam line}] (v1f);
		\coordinate[right = of v1f] (v1g);
		\coordinate[right = of v1g] (v1h);
		\coordinate[right = of v1h] (v1i);
		\coordinate[right = of v1i] (v1l);
		\coordinate[right = of v1l] (v1m);
		\coordinate[right = of v1m] (v2a);
		\coordinate[right = of v2a] (v2b);
		\coordinate[right = of v2b] (v2c);
		\coordinate[right = of v2c] (v2d);
		\coordinate[right = of v2d] (v2e);
		\coordinate[right = of v2e] (v2f);
		\coordinate[right = of v2f] (v2i);
		\coordinate[right = of v2i] (v2j);
		\coordinate[right = of v2j] (v2k);
		\coordinate[right = of v2k] (v2l);
		\coordinate[right = of v2l] (v2m);
		\coordinate[right = of v2m] (v2n);
		\coordinate[right = of v2n] (v2);
		\coordinate[below = of v1a] (v3b);
		\coordinate[below = of v3b] (v3c);
		\coordinate[below = of v3c] (v3d);
		\coordinate[below = of v3d] (v3e);
		\coordinate[below = of v3e] (v3f);
		\coordinate[below = of v3f] (v3g);
		\coordinate[below = of v3g] (v3h);
		\coordinate[below = of v3h] (v3);
		\coordinate[above = of v2n] (v4a);
		\coordinate[above = of v4a] (v4b);
		\coordinate[above = of v4b] (v4c);
		\coordinate[above = of v4c] (v4d);
		\coordinate[above = of v4d] (v4e);
		\coordinate[above = of v4e] (v4f);
		\coordinate[above = of v4f] (v4g);
		\coordinate[above = of v4g] (v4);
		\draw[fermionnoarrow] (v1) -- (v2);
		\draw[thick,red] (v3) rectangle (v4);
		\draw[fill] (v2a) circle [radius=2pt];
		\coordinate[above = of v2i] (v5a);
		\coordinate[above = of v5a] (v5);
		\draw[fill] (v5) circle [radius=1pt];
		\draw[fermionnoarrow] (v2a) -- (v5);
		\coordinate[above right = of v5] (v6a);
		\coordinate[above right = of v6a] (v6b);
		\coordinate[above = of v6b] (v6c);
		\coordinate[above = of v6c] (v6d);
		\coordinate[above = of v6d] (v6e);
		\coordinate[above = of v6e] (v6);
		\draw[fermionnoarrow] (v5) -- (v6);
	\end{tikzpicture}
	\caption{Process where a long-lived highly charged state decays to a collider-stable singly charged state, leading to events where activity in the muon chambers is not aligned with the charged track.  (Left): view along the beam pipe.  (Right): transverse view.  The red lines show the edges of the detector.  As discussed in the text, these events are difficult to identify and lead to a weakening of limits around $\Delta m_1 \sim 20$--30\;MeV.}\label{fig:tricky}
\end{figure}
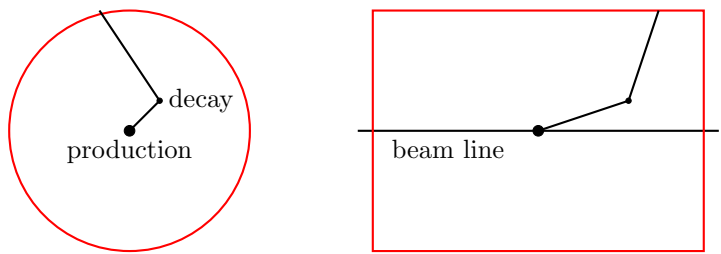

For the quintets and septets, similar conclusions apply to the extension of the monophoton searches to the maximal mass splitting.  However, as alluded to in \secref{sec:llcp}, there are potential problems that arise for smaller mass splittings, $\Delta m_1 \sim 20$--30\;MeV.  At this point, the doubly- and triply-charged states in the multiplet become long-lived themselves, which can complicate the experimental signals.   If all the charged states either decay promptly or are collider-stable, then there are no problems with applying the LLCP limits.  The difficulty comes when one (or more) states travel into the detector, but decay to a collider-stable particle \emph{before} the muon chambers.  These events are sketched in \figref{fig:tricky}.  The resultant break in the particle track will interfere with measuring the LLCP velocity and identifying it, weakening the bounds.

The worst case scenario would be if all production of the doubly and/or triply charged states would lead to signals of this kind, \emph{and} that these events could not be identified as arising from BSM physics instead of \emph{e.g.} cosmic rays.  Even in this case, we still have limits from direct production of the collider-stable singly charged state.  The production cross section of this specific state is the same no matter the multiplet, so \emph{at worst} the constraints on the quintet and septet will be the same as for the triplet.  For fermionic models, the triplet limits from \figref{fig:combFerm} extend to the kinematic limit, $m_\chi > 1.5$\;TeV, in the relevant mass range.  The results for the large fermion multiplets are unchanged.  In contrast the constraint on the scalar triplet from \figref{fig:combSca} is only $m_\chi \gtrsim 1.425$\;TeV, and there is a potential weakening of the bounds.  We leave a more precise study of this effect to future work.

%% file: Files/Pions.tex
The most obvious feature from the combined plots of the previous section is the relative weakness of the monophoton searches compared to LLCPs and disappearing tracks, and the correspondingly unconstrained regions of parameter space at moderately large mass splittings $\Delta m_1 \gtrsim 200$\;MeV.  This is a consequence of the signal and irreducible background having identical polarisation dependence, removing one of the main handles usually used to improve the signal-to-background ratio.  In this section we discuss some alternative strategies that might provide greater sensitivity.

\begin{figure}
	\centering
	\begin{tikzpicture}[node distance=0.75cm and 0.75cm]
		\coordinate (v1);
		\coordinate[below = of v1] (v2);
		\coordinate[right = of v1, label=above: $W^-$] (v3);
		\coordinate[right = of v2, label=below: $W^+$] (v4);
		\coordinate[right = of v3] (v5);
		\coordinate[right = of v4] (v6);
		\coordinate[above left = of v1, label=above left:$e^-$] (i1);
		\coordinate[below left = of v2, label=below left:$e^+$] (i2);
		\coordinate[above right = of v5, label=above right:{$\psi^{-}$}] (o1);
		\coordinate[right = of v5, label=right:{$\chi$}] (o2);
		\coordinate[right = of v6, label=right:{$l^{+}$}] (o3);
		\coordinate[below right = of v6, label=below right:{$\nu_l$}] (o4);
		\draw[fermion] (i1) -- (v1);
		\draw[fermion] (v1) -- (v2);
		\draw[fermion] (v2) -- (i2);
		\draw[photon] (v1) -- (v5);
		\draw[photon] (v2) -- (v6);
		\draw[fermionnoarrow] (o2) -- (v5) -- (o1);
		\draw[fermion] (o3) -- (v6);
		\draw[fermion] (v6) -- (o4);
	\end{tikzpicture}\qquad\qquad
	\begin{tikzpicture}[node distance=0.75cm and 0.75cm]
		\coordinate (v1);
		\coordinate[right = of v1] (v15);
		\coordinate[right = of v15] (v2);
		\coordinate[below = of v2, label=above left:$W^+$] (v3);
		\coordinate[above left = of v1, label=above left:$e^+$] (i1);
		\coordinate[below left = of v1, label=below left:$\gamma$] (i2);
		\coordinate[above right = of v2, label=above right:{$\nu_e$}] (o1);
		\coordinate[above right = of v3, label=above right:{$\nu_l$}] (o2);
		\coordinate[below right = of v3, label=below right:{$l^+$}] (o3);
		\draw[fermion] (v1) -- (i1);
		\draw[photon] (v1) -- (i2);
		\draw[fermion] (v2) -- (v1);
		\draw[fermion] (o1) -- (v2);
		\draw[photon] (v2) -- (v3);
		\draw[fermionnoarrow] (o2) -- (v3);
		\draw[fermionnoarrow] (v3) -- (o3);
	\end{tikzpicture}
	\caption{Illustrative processes for potential mono-lepton searches.  (Left): Signal process $e^+e^-\to \psi^+\chi l^- \nu_l$ for lepton $l$.  (Right): Beamsstrahlung-initiated irreducible background relevant for all leptons.}\label{fig:monol}
\end{figure}
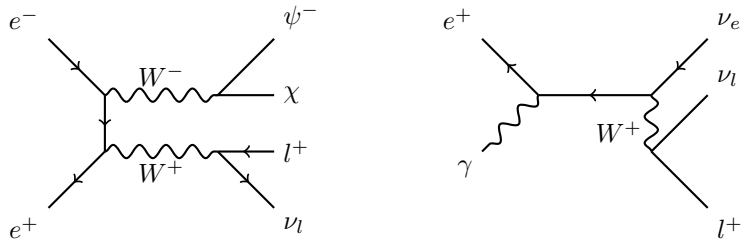

The simplest possibilities to consider are other searches based on a single energetic object and no other detector activity.  When the charged state decay products are invisible, we can produce final states with no $e^+e^-$-initiated irreducible background.  For example, $e^+e^-\to \psi^\mp\chi l^\pm \nu$ with $l = e, \mu$ will appear as a single charged lepton recoiling against nothing, see the left of \figref{fig:monol}.  Since $e^+e^-$-initiated SM processes have zero net charge, they can only contribute to the background when soft or collinear objects are not reconstructed.  Unfortunately, this search suffers from the \emph{beamsstrahlung-initiated} irreducible background $e^\pm\gamma \to l^\pm\nu\nu$, see the right of \figref{fig:monol}.  This has fewer final state particles than the signal, which compensates for arising through radiative effects; and with both arising through weak interactions, polarisation is again of little use in enhancing the signal.  Together with the smaller signal cross section, the prospects are worse than for monophotons.  Similar problems arise for hadronic $W$ decays.

The fundamental weakness of these searches is that they do not exploit all the information in the event.  Our dark sector charged particles \emph{do} leave visible signals through their decay products.  Previous work has used the very soft decay products of Higgsinos to set prospective limits at the ILC~\cite{1307.3566}.  However, we so far neglected them because they are quite challenging to observe: not only very soft, but superimposed upon a substantial $\gamma\gamma \to$\;hadrons background of additional soft charged tracks.  Even if the decay products' tracks can be cleanly reconstructed, we need some means to identify their origin \emph{as} the decay of the charged members of the DM multiplet.  This guides us to once again consider $\psi$ production in association with energetic SM particles.  Specifically we need to consider \emph{charged} visible matter, so that we can use the energetic track(s) to identify the primary interaction vertex.  Soft tracks emerging from the same position are then candidates for these decay products.

The two most promising signals are then one or two energetic leptons, together with coincident soft tracks.  The backgrounds consist of hard SM events with the same final states that happen to be spatially coincident with a soft $\gamma\gamma \to$\;hadrons event.  Both of the signal processes are four-particle final states, so we expect them to have similar cross sections.  The one-lepton process has the advantage of no $e^+e^-$-initiated irreducible contribution to the hard component of the background.  However, for the doublet and triplet models, where improvement over monophotons is most urgently needed, this channel only produces a single dark sector decay and so typically a single soft track.  In contrast, the process $e^+e^- \to \psi^+\psi^- l^+l^-$ will involve at least two soft tracks, reducing the background fake rate and potentially compensating for the larger hard background.

The potential reach in this channel is naturally sensitive to the details of $\gamma\gamma \to$\;hadrons in the CLIC environment, as well as the detector response to very soft objects.  Since we do not have reliable information on these topics, we will not attempt to compute the discovery reach in detail, deferring it to future work.  We instead restrict ourselves to some qualitative observations about the potential limits and their shape in the mass-lifetime plane.  To understand both, it is useful to consider how the signal depends on the mass splitting $\Delta m_1$, which it does in two distinct ways.  The more obvious arises from requiring the hard and soft tracks to reconstruct a common vertex.  The strongest background rejection is obtained for demanding the lines meet to the measurement accuracy, and for CLIC the design goal is $\mathcal{O}(10\;\mu$m)~\cite{1202.5940}.  The maximum signal acceptance will occur for lifetimes below this scale.  From \figrefs{fig:combFerm} and~\ref{fig:combSca}, we see that we have interest in larger lifetimes; in these cases, it might be better to relax the vertex reconstruction criteria to increase acceptance at the cost of larger backgrounds.  The second manner in which $\Delta m_1$ influences the signal is that the decay products will have $p_T \sim \gamma \Delta m_1$, where $\gamma$ is the boost of the parents.  Since $\gamma \sim \sqrt{s}/2m_\chi$, we can see how even very modest GeV-scale cuts on the track $p_T$ will heavily reduce the signal.

Our signal process is defined by three main cuts: a cut $p_T > p_T^e$ on the hard object(s), a cut $p_T > p_T^{soft}$ on the soft tracks, and a vertex size $d$.  For a candidate choice $p_T^e = 100$\;GeV, the irreducible hard background cross sections are $\sim 1$\;pb (5\;pb) for the one lepton (two muon) events.  This is temporally coincident with approximately 20 soft interactions, arising from multiple bunch interactions.  With a bunch spacing of 0.5\;ns, we can estimate a longitudinal size for the interaction region of $\mathcal{O}(10$\;cm), comparable to that of the LHC.  The probability of one of these events overlapping with a SM hard event to the precision $d$ is given by
\begin{equation}
	\mathcal{P}^{fake} \approx 20 \times \frac{d}{10\;\text{cm}} = 0.2\% \, \frac{d}{10\,\mu\text{m}}.
\label{eq:pfake}\end{equation}
This leads to a background cross section of at least 2\;fb (10\;fb) for our two signals, and likely more thanks to the contribution of reducible processes.  The hard background processes are pure electrodynamics, and as such a systematic uncertainty at the percent level is a reasonable expectation.  The $\gamma\gamma \to$\;hadrons contribution is more difficult, but can be extracted from data by comparing the activity in the mono-electron events as here with mono-photon events.  If we assume a 10\% systematic uncertainty then we have a potential exclusion $\sigma^{sig} \lesssim 0.4$\;fb (2\;fb).  For a fermion triplet, this cross section (before cuts) corresponds to a mass of $\sim 1150$\;GeV (700\;GeV).  We emphasise that this is an optimistic estimate of the reach.

The strongest constraint will apply to the maximal mass splitting, $\Delta m_1 \sim 1$\;GeV.  As the lifetime increases, the two effects discussed above will weaken the limits.  We consider first the effect of $p_T^{soft}$.  If it is possible to use $p_T^{soft} = 1$\;GeV,\footnote{CLIC has a design goal of a 99\% reconstruction efficiency for tracks at least this hard.}, then nearly all events at maximal mass splitting will pass this cut.  At smaller mass splittings, the need for the decay products to be boosted gives a maximum mass reach of $\sim (\Delta m_1/$GeV)$\times 1.5$\;TeV.  This is greater than the optimal fermion triplet limits above for $\Delta m_1 \gtrsim 770$\;MeV (470\;MeV); these large mass splittings are unaffected by this cut.  At smaller mass splittings the reach decreases, becoming worse than the monophoton bound for $\Delta m_1 \lesssim 300$\;MeV.

The lifetime constraint is likely a more important effect.  Based on our estimates above, we are systematics-dominated, $\epsilon N_{bkg} \gg \sqrt{N_{bkg}}$.  From \modeqref{eq:sig}, the limit on the signal cross section is then proportional to the background cross section, and thus $\mathcal{P}^{fake}$ and finally $d$.  If we choose to maximise our signal acceptance by taking $d = c\tau$, and additionally make the na\"\i ve estimate $\sigma \sim m_\chi^{-2}$, we can estimate that the excluded mass would be roughly proportional to $(c\tau)^{-1/2}$.  The reach would drop below 100\;GeV for $c\tau \gtrsim 10^{-3}$\;m, or a triplet mass splitting of $\Delta m_1 \lesssim 500$\;MeV.  A better sensitivity might arise from smaller choices of $d$, but this is beyond our analysis here.

%% file: Files/conc.tex
In this work, we have examined how the proposed $e^+e^-$ collider CLIC can constrain and discover electroweak multiplet dark matter through direct searches.  The models we consider are difficult to test with direct detection experiments as the DM itself has no (for fermions) or only very weak (for scalars) elastic couplings with the SM.  Cosmic ray searches are stronger but face unavoidable systematic uncertainties.  Collider searches, in contrast, can make unambiguous statements about the presence or absence of matter coupling through SM gauge interactions.  In theories without any light coloured states, lepton colliders are an efficient search tool as production is efficient and they have simpler detector environments.

The absence of a $\chi$-SM coupling implies that only the charged members of the multiplets are produced at CLIC.  The phenomenology is almost entirely determined by the lifetime of the singly charged state $\psi^+$, which in turn is determined by $\Delta m_1 = m_{\psi^+} - m_\chi$.  We discussed the origin of this mass splitting in \secref{sec:models}, and in particular noted that for all models other than the fermion doublet $\Delta m_1 \lesssim 1$\;GeV.  Despite this relatively small range, the lifetime varies over several orders of magnitude due to the presence of multiple decay thresholds, especially the muon and pion ones at $\Delta m_1 \sim 100$\;MeV.  This leads to several distinct signals:
\begin{itemize}
	\item At mass splittings well below the $\mu/\pi$ mass, $\psi^+$ is collider stable, $c\tau > 10$\;m.  The strongest limits come from searches for long-lived charged particles discussed in \secref{sec:llcp}; for most multiplets, we can easily exclude up to the kinematic limit $m_\chi = 1.5$\;TeV.  As the lifetime decreases, these searches fail due to $\psi^+$ decaying within the detector.
	\item At larger mass splittings close to the muon and pion mass, $\psi^+$ can travel macroscopic distances but still decay within the detector volume.  This leads to a `disappearing track' signal consisting of a hard charged track, little calorimeter activity, and nothing in the muon chambers; we discuss this in \secref{sec:dtrack}.  Because we invoke the presence of a hard additional photon to eliminate fakes, these searches can test at best $m_\chi \approx 1.4$\;TeV.  The smaller production cross section of scalar multiplets means the reach there is weaker.  These searches fail at large lifetimes when $\psi^+$ enters the muon chamber, and at short lifetimes when it decays within the beam pipe.
	\item  Finally, well above the muon/pion thresholds $\psi^+$ decays promptly.  The soft nature of its SM decay products make them essentially invisible against the $\gamma\gamma\to$\;hadrons background.  The strongest limits we found derive from monophotons, where the DM recoils against an energetic photon, and were studied in \secref{sec:monophotons}.  Because of the large irreducible backgrounds and the signal and background having the same dependence on the beam polarisation, these limits are below 1\;TeV except for the fermion quintet and septet.
\end{itemize}
We combine all these results in \secref{sec:comb}, plotting the potential discovery and exclusion contours in the mass-lifetime plane for fermions in \figref{fig:combFerm} and for scalars in \figref{fig:combSca}.  When the mass splitting is set purely by radiative effects, the reach is given in \tabref{tab:radexc}.  In particular, we can test the Higgsino-like doublet to 310\;GeV; the wino to 775\;GeV; and minimal dark matter to 775\;GeV.  For the Higgsino-like model, we can test the thermal relic mass $m_\chi = 1$\;TeV for mass splittings $\Delta m_1 < 230$\;MeV.

The obvious weakness in the searches we have considered is the relative insensitivity of monophoton searches.  Only the two larger fermion models can be tested this way for masses over 1\;TeV.  Importantly, note that for all models other than the fermion and scalar triplets, monophotons place the strongest bounds when the mass splitting arises only from radiative corrections.  Accordingly, we discussed possible avenues of improvement in \secref{sec:pions}.  The most promising strategy would seem to be exploiting all of the final state information by attempting to identify the soft decay products of the dark matter.  This is challenging due to the large coincident $\gamma\gamma \to$~hadrons background, so we propose using one or two additional hard charged tracks as a tool to identify the primary vertex.  A full calculation of the reach would require a better understanding of the background than is currently available, but we gave estimates of what the possible reach might be for the fermion triplet.  Under ideal conditions they could test twice the mass accessible via mono-photons, with the best sensitivity achievable at large values of $\Delta m_1$.

The nature of DM will be a key question any future collider will hope to illuminate.  In this work we have shown that linear colliders can be powerful tools in exploring a class of models that are simple but also motivated by top-down theories.  Direct searches are limited to the centre of mass energy, but are also insensitive to the kinematically inaccessible states except through the mass splitting.  The results we have found are then robust, conservative and unequivocal statements that CLIC, or any similar linear collider, can make about the presence or absence of these stable electroweak multiplets.